\documentclass[english]{revtex4-2}
\usepackage[latin9]{inputenc}
\usepackage{color}
\usepackage{babel}
\usepackage{array}
\usepackage{multirow}
\usepackage{amsmath}
\usepackage{graphicx}
\usepackage{rotating}
\usepackage[unicode=true,pdfusetitle,
 bookmarks=true,bookmarksnumbered=false,bookmarksopen=false,
 breaklinks=false,pdfborder={0 0 1},backref=false,colorlinks=true]
 {hyperref}

\makeatletter

\providecommand{\tabularnewline}{\\}

\makeatother

\begin{document}
\title{Excited muon production at muon colliders via contact interaction}
\author{M. Sahin}
\email{mehmet.sahin@usak.edu.tr}

\affiliation{Faculty of Sciences and Literature, Department of Physics, Usak University,
64200, Usak, Turkey}
\author{A. Caliskan}
\email{acaliskan@gumushane.edu.tr}

\affiliation{Faculty of Engineering and Natural Sciences, Department of Physics
Engineering, Gümü\c{s}hane University, 29100, Gümü\c{s}hane, Turkey }
\begin{abstract}
In recent years, with the enlightenment of some issues encountered
at muon colliders, muon colliders have become more feasible for the
high-energy physics community. For this reason, we studied single
production of excited muon at muon colliders via contact interaction.
Besides, we assumed that the excited muon is produced via contact
interactions and decays to the photon and the muon through the gauge
interaction. Then, signal and background analyses were performed at
the muon anti-muon collider options with 6 TeV, 14 TeV, and 100 TeV center-of-mass energies for the excited muon. Attainable mass and compositeness
scale limits were calculated for the excited muon at the muon anti-muon colliders.
As a result of the calculations, it was concluded that the muon-antimuon colliders would be a perfect collider option for the excited muon investigations.
\end{abstract}

\maketitle

\section*{1. INTRODUCTION}

The Standard Model is a theory that explains the best manner of basic
building blocks of the universe. It describes fermions make up all
the matter that interacts with light that makes up all stars and galaxies.
The Standard Model also describes force-carrying particles that are
called bosons. It also to explains how elementary particles gain mass
with the discovery of the Higgs Boson, the last missing part of the
Standard Model. Although the Standard Model successfully explains
the nature of the universe, it does not answer some questions. As
an example of these questions, why does the Standard Model have many
particles and their parameters? Many theories have been developed
beyond the standard model to answer these questions. Composite models,
string theory, extra dimensions, and supersymmetry are the most known
beyond the Standard Models.

The number of particles and parameters in the Standard Model has been
reduced by composite models in the best manner \citep{Pati:1974yy,Terazawa:1976xx,Shupe:1979fv,Harari:1979gi,Terazawa:1979pj,Fritzsch:1981zh,Terazawa:1981eg,Lyons:1982jb,Terazawa:1983yn,Eichten:1983hw,DSouza:1992ljp,Celikel:1998dj,de2008weak,terazawa2014composite,Terazawa:2015bsa,Fritzsch:2016jpg,kaya2018minimal}.
Based on the composite models, the elementary particles in Standard Models
have an internal substructure. Since these elementary particles have
a composite structure, they are composed of more fundamental particles
called preons. If excites states of the Standard Model quarks and
leptons are discovered by particle colliders, this discovery revealed
that quarks and leptons have a compound substructure. Excited states
of the Standard Model quarks and leptons are called excited quarks
and excited leptons or excited fermion in particle physics. Excited fermions possess spin-1/2 and spin-3/2 quantum numbers. There are many
publications on the excited fermions in the literature \citep{Low:1965ka,renard1983excited,Kuhn:1984rj,Pancheri:1984sm,DeRujula:1983ak,hagiwara1985excited,Kuhn:1985mi,baur1987excited,spira1989excited,Baur:1989kv,Jikia:1989gj,boudjema1993excited,Cakir:1999nu,Cakir:2000vt,Cakir:2000sw,Eboli:2001hi,Cakir:2004qd,Cakir:2002eu,Cakir:2004cv,Cakir:2007wn,Cakir:2008hw,Ozansoy:2012vb,Koksal:2014gca,Biondini:2014dfa,Ozansoy:2016ivj,Panella:2017spx,Caliskan:2017meb,Caliskan:2017fts,Gunaydin:2017enp,Caliskan:2018vsk,Akay:2018ezz,Sahin:2019nqy,Biondini:2019tcc,Caliskan:2018pdq,ccalicskan2018single,Buskulic:1996tw,Abreu:1998jw,Abbiendi:1999sa,Achard:2003hd,Aaron:2008cy,Acosta:2004ri,Abulencia:2006hj,Abazov:2006wq,Abazov:2008ag,Aad:2013jja,Aad:2016giq,Aaboud:2019zpc,CMS:2012ad,Khachatryan:2015scf,Sirunyan:2018zzr,Sirunyan:2020awe,10.1093/ptep/ptaa104}.
This indication shows that the issues related to excited fermions
are the broad interest to particle physicists. Experimental studies
on the excited leptons at the CERN LEP \citep{Buskulic:1996tw,Abreu:1998jw,Abbiendi:1999sa,Achard:2003hd},
DESY HERA \citep{Aaron:2008cy}, Fermilab Tevatron \citep{Acosta:2004ri,Abulencia:2006hj,Abazov:2006wq,Abazov:2008ag}
have found no evidence for the excited leptons. Also, ATLAS \citep{Aad:2013jja,Aad:2016giq,Aaboud:2019zpc}
and CMS \citep{CMS:2012ad,Khachatryan:2015scf,Sirunyan:2018zzr,Sirunyan:2020awe}
Experiments found no hints of excited leptons at the Large Hadron Collider. By assuming the mass of the excited lepton
equal to the compositeness scale, the CMS experiment excluded excited
electrons and muons with masses below 3.8 and 3.9 TeV, respectively,
in the channel where they decay to two leptons and one photon through
gauge mode decay \citep{Sirunyan:2018zzr}. In the same analysis,
the CMS experiment excluded the compositeness scale up to 25 TeV for
$M_{l^{*}}=1$ TeV. In the next study, the CMS experiment excluded
excited electrons and muons with masses below 5.6 and 5.8 TeV, respectively,
for $M_{l^{*}}=\Lambda$ \citep{Sirunyan:2020awe}. In this study
of the CMS experiment, the excited electron and the muon decayed to
one lepton and two jets through contact interaction.

If quarks and leptons consist of more fundamental sub-components,
new interactions should arise between quarks and leptons at the compositeness scale energies of the SM fermions. Besides, if
the collider's center-of-mass energy is much lower than the compositeness
scale, these new types of interactions are suppressed by the inverse
powers of $\varLambda$. In this case, four fermion contact interactions
are the best way to investigate fermions' compositeness and excited
fermions. In this study, we have investigated spin-1/2 excited muon
production via contact interaction at muon colliders. Excited muons
decay to muon and photon through electromagnetic interactions. In
the second section, we give knowledge about the TeV scale muon colliders.
We present contact and gauge interaction lagrangian of the excited muons
as well as contact and gauge decay widths and signal cross-sections
for the process of $\mu^{-}\mu^{+}\rightarrow\mu^{*-}\mu^{+}\rightarrow\gamma\mu^{-}\mu^{+}$
in the third section. In the fourth chapter, signal and background
analysis at the muon colliders are presented. Lastly, we present an explanation of obtained results.

\section*{2. TEV SCALE MUON COLLIDERS}

According to the standard model, the leptons are elementary particles,
and the lepton colliders are more advantageous than proton colliders.
Only some of the collision's energy can produce new particles in the
proton colliders since the protons are composed of quarks. On the
other hand, the lepton colliders offer point-like collisions, and
all of the energy from the collision is used for particle production
\citep{ryne:muon2020}

The biggest problem with circular electron accelerators is synchrotron
radiation, which occurs in bending magnet regions. Since electrons
are bent while passing through bending magnets, they lose most of
their energy by producing synchrotron radiation. This loss of energy
restricts the electrons from being accelerated to higher energies
in circular accelerators. Due to this technical challenge, it is planned
to use long linear accelerators to reach more high energies in international
electron-positron collider projects, planned to be established in
the future, such as ILC \citep{behnke:international2013,baer_international_2013,adolphsen_international_2013,adolphsen_international_2013-1,behnke_international_2013-1}
and CLIC \citep{aicheler_multi-tev_2012,linssen_physics_2012,lebrun_clic_2012}.
It is the case that the fact that muon is $207$ times heavier
than the electron, making the muon collider more attractive. The problem
of synchrotron radiation is suppressed mainly in the muon colliders.
Heavy muons can be accelerated to high energies without significant
power losses in smaller circular colliders. The muon-antimuon colliders
are expected to be the most efficient device for new physics researches
\citep{long_muon_2021}.

The international muon collider project, which uses a proton-driven
source, was first launched in the United States in 2011 under the
name of the Muon Accelerator Program (MAP) \citep{MuonAcceleratorProgram}.
In this project, three muon collision options are proposed. The collision
center-of-mass energies of these options are $1.5$, $3$, $6$ TeV
and their luminosity values are $1.25$, $4.4$ and $12\times10^{34}cm^{-2}s^{-1}$,
respectively. In this study, the collision option of $6$ TeV is taken
into account. Other technical information and parameters for proton-driven
muon colliders can be found in \citep{boscolo2019future} and \citep{delahaye2019muon}.

In the proton-driven muon collider scheme, the muon beam's production
and conversion into the bunched structure is a complicated process.
The proton beam is struck at a target to produce secondary $\pi$-mesons.
These unstable particles immediately decay into muons and neutrinos.
The particles in the generated muon beam have a wide spread at different
positions and velocities. In order to convert the muon beam into a
bunched structure, it is necessary to reduce this spread, that is,
reduce the phase space \citep{ryne:muon2020,long_muon_2021}. This
process, which is done to obtain a high-quality bunched muon beam,
is called beam cooling. Traditional beam cooling techniques used experimentally,
such as laser cooling \citep{schroder_first_1990} and stochastic
cooling \citep{mohl_physics_1980}, are not suitable for the muon
beam. None of these methods are fast enough to cool them because the
muons are unstable and short-lived particles. An ionization cooling technique
was proposed to cool the muon beam \citep{neuffer_principles_1983},
and various theoretical and numerical studies on this method were
carried out. However, it was not experimentally proven until 2019.
The critical point in realizing proton-driven muon colliders depended
on the efficient application of this technique. The proof of the ionization
cooling technique was finally realized by the MICE (Muon Ionization
Cooling Experiment) collaboration for the first time at the Rutherford
Appleton Laboratory in the UK \citep{mice_collaboration_demonstration_2020}.
However, not a $100$ \% efficient experiment, demonstrating this
technique is an essential step towards obtaining a quality muon beam.
This work of MICE collaboration is a milestone in the realization
of the proton-driven muon colliders.

A different international project has recently been proposed for the
muon collider, which is called the Low EMittance Muon Accelerator
(LEMMA) program \citep{antonelli_novel_2016}. In this project, the
pair production of the muons is carried out after the interaction
of high-intensity positron beams with electrons at a fixed target
according to the $e^{+}e^{-}\rightarrow\mu^{+}\mu^{-}$ annihilation
process. This technique's advantage is that the muons produced have
a very small emittance value, and therefore no cooling process is
required. Experimental testing of this method, which has technical
difficulties such as low muon production rate, is still ongoing \citep{noauthor_lemma_2020,bartosik_muon_2019}.
A center-of-mass collision energy of $6$ TeV with a luminosity of
$5.1\times10^{34}cm^{-2}s^{-1}$ has been proposed for this positron-driven
muon collider project \citep{boscolo2019future}.

Some feasibility studies have been recently carried out to upgrade
the LHC and FCC complexes to a muon collider \citep{neuffer_feasibility_2018,zimmermann_lhc/fcc-based_2018}.
Three options have been offered for the proposed $14$ TeV muon collider
project installed in the LHC tunnel. In the first option, called the
PS option, a $24$ GeV energy PS (Proton Synchrotron) source available
at CERN will be used, while in the second, an $8$ GeV energy linac
and subsequently a storage ring will be required (MAP option). It is planned to use the proton-driven muon source technology proposed in the MAP project for these two options. In the third option, called
the low emittance muon collider (LEMC) option, positron-driven muon
source technology will be used. The luminosity values of the three
muon collider options suggested to be installed in the LHC tunnel
are $1.2\times10^{33}$, $3.3\times10^{35}$, and $2.4\times10^{32}cm^{-2}s^{-1}$,
respectively \citep{neuffer_feasibility_2018}. In this study, the
first (proton-driven muon source) luminosity values and third (positron-driven
muon source) options, which are easier to reach experimentally, were
used.

The FCC complex \citep{fcc2019fcc,fcc2019fcc:ee,fcc2019fcc:hh,fcc2019he}
is planned to be established in the future, and it can also be converted
into a $100$ TeV muon collider. Feasibility studies for various options
related to this collider are still ongoing. A luminosity value of
$1\times10^{34}cm^{-2}s^{-1}$ was used in this study for the muon
collider of $100$ TeV \citep{zimmermann_lhc/fcc-based_2018}. Table
1 shows the center-of-mass energies and luminosity values of the muon
colliders used in this study.
\begin{widetext}
\begin{table}
\caption{\label{tab:Muon-antimuon-colliders-and}Muon-antimuon colliders and
their main parameters}
\begin{tabular}{|c|c|c|c|c|c|}
\hline 
\multirow{2}{*}{} & \multirow{2}{*}{MAP} & \multirow{2}{*}{LEMMA} & \multicolumn{2}{c|}{LHC-$\mu\mu$} & \multirow{2}{*}{FCC-$\mu\mu$}\tabularnewline
\cline{4-5} \cline{5-5} 
 &  &  & PS option & LEMC option & \tabularnewline
\hline 
Center-of-Mass Energy {[}TeV{]} & $6$ & $6$ & $14$ & $14$ & $100$\tabularnewline
\hline 
Luminosity {[}$cm^{-2}s^{-1}${]} & $12\times10^{34}$ & $5.1\times10^{34}$ & $1.2\times10^{33}$ & $2.4\times10^{32}$ & $1\times10^{34}$\tabularnewline
\hline 
\end{tabular}
\end{table}
\end{widetext}

\section*{3. INTERACTION LAGRANGIAN, DECAY WIDTHS AND, CROSS SECTIONS}
\begin{widetext}
The Standard Model fermions possess three families. It is thought
that the excited fermions will have three families as similar Standard
Model fermion. When left-handed and right-handed components of excited
leptons and quarks are assigned to the isodoublet spin structure,
the isospin structures of three-family excited leptons, excited quarks,
and Standard Model fermions can be represented by Equation \ref{1,eq:wideeq1}.

\begin{eqnarray*}
\left[\begin{array}{c}
\nu_{e}\\
e^{-}
\end{array}\right]_{L},e_{R}^{-},\left[\begin{array}{c}
\nu_{e}^{*}\\
e^{*-}
\end{array}\right]_{L},\left[\begin{array}{c}
\nu_{e}^{*}\\
e^{*-}
\end{array}\right]_{R},\left[\begin{array}{c}
u\\
d
\end{array}\right]_{L},u_{R},d_{R},\left[\begin{array}{c}
u^{*}\\
d^{*}
\end{array}\right]_{L},\left[\begin{array}{c}
u^{*}\\
d^{*}
\end{array}\right]_{R} & \;
\end{eqnarray*}

\begin{eqnarray}
\left[\begin{array}{c}
\nu_{\nu}\\
\mu^{-}
\end{array}\right]_{L},\mu_{R}^{-},\left[\begin{array}{c}
\nu_{\mu}^{*}\\
\mu^{*-}
\end{array}\right]_{L},\left[\begin{array}{c}
\nu_{\mu}^{*}\\
\mu^{*-}
\end{array}\right]_{R},\left[\begin{array}{c}
c\\
s
\end{array}\right]_{L},c_{R},s_{R},\left[\begin{array}{c}
c^{*}\\
s^{*}
\end{array}\right]_{L},\left[\begin{array}{c}
c^{*}\\
s^{*}
\end{array}\right]_{R} & \;\label{1,eq:wideeq1}
\end{eqnarray}

\begin{eqnarray*}
\left[\begin{array}{c}
\nu_{\tau}\\
\tau^{-}
\end{array}\right]_{L},\tau_{R}^{-},\left[\begin{array}{c}
\nu_{\tau}^{*}\\
\tau^{*-}
\end{array}\right]_{L},\left[\begin{array}{c}
\nu_{\tau}^{*}\\
\tau^{*-}
\end{array}\right]_{R},\left[\begin{array}{c}
t\\
b
\end{array}\right]_{L},t_{R},b_{R},\left[\begin{array}{c}
t^{*}\\
b^{*}
\end{array}\right]_{L},\left[\begin{array}{c}
t^{*}\\
b^{*}
\end{array}\right]_{R} & \;
\end{eqnarray*}
\end{widetext}

Excited leptons could also be produced via the contact interaction
with Standard Model quarks and leptons at particle colliders. When
the energy of the particle collider is below the compositeness scale,
four-fermion contact interactions become significant. Thus, we define
these interactions as in Equation \ref{eq:2} \citep{Baur:1989kv,10.1093/ptep/ptaa104}.
\begin{widetext}
\begin{eqnarray}
L_{CI}=\frac{g_{*}^{2}}{\Lambda^{2}}\frac{1}{2}j^{\mu}j_{\mu}\label{eq:2}
\end{eqnarray}

\begin{equation}
j_{\mu}=\eta_{L}\overline{f_{L}}\gamma_{\mu}f_{L}+\eta_{L}^{'}\overline{f_{L}^{*}}\gamma_{\mu}f_{L}^{*}+\eta_{L}^{''}\overline{f_{L}^{*}}\gamma_{\mu}f_{L}+H.c.+(L\rightarrow R)\;.\label{eq:wideeq2}
\end{equation}
\end{widetext}

Where $g_{*}$ is contact interaction coupling and their value is
taken as $g_{*}^{2}=4\pi.$ $j_{\mu}$ called left-handed currents
and, $\eta_{L}$, $\eta_{L}^{'}$, $\eta_{L}^{''}$ are coefficients
of left-handed currents. Their values are equal to one in this study.
Besides, right-handed helicity currents are omitted for clarity and brevity. $\Lambda$ represents the compositeness scale. 

Also, excited leptons can decay standard model quarks and leptons
through contact interactions, as well as decay standard model quarks
and leptons through gauge interactions. The Standard Model gauge interactions
of excited leptons can be represented by Equation \ref{eq:4} \citep{hagiwara1985excited,Baur:1989kv,10.1093/ptep/ptaa104}.

\begin{eqnarray}
L_{G}=\frac{1}{2\Lambda}\overline{l_{R}^{*}}\sigma^{\mu\nu}\left(gf\frac{\overrightarrow{\tau}}{2}\overrightarrow{W_{\mu\nu}}+f^{'}g^{'}\frac{Y}{2}B_{\mu\nu}\right)l_{L}+H.c.\label{eq:4}
\end{eqnarray}

Here $l_{L}$ represents the left-handed Standard Model lepton doublet,
$l_{R}^{*}$ denotes the right-handed excited lepton doublet and,
$\overrightarrow{W_{\mu\nu}}$ and $B_{\mu\nu}$ are field strength
tensor in the Equation \ref{eq:4}. $Y$ and $\overrightarrow{\tau}$
are weak hyper-charge and Pauli spin matrices, respectively. Gauge
coupling constant are $g$ and $g^{'}$. Free parameters are $f$
and $f^{'}$ determined by the dynamics of compositeness. We assume
that their value is $f=f^{'}=1$. Excited leptons decay Standard Models
leptons via gauge interaction or contact interactions. The analytic
formulas of the excited leptons decay to the Standard Model leptons via
gauge interactions are described below. Firstly, we present the excited
lepton's partial decay width through photon emission in Equation \ref{eq:5}. 

\begin{eqnarray}
\varGamma\left(l^{*}\rightarrow\gamma l\right)=\frac{1}{4}\alpha f_{\gamma}^{2}\frac{m_{l^{*}}^{3}}{\Lambda^{2}}\label{eq:5}
\end{eqnarray}

Secondly, we introduce the partial decay width of excited lepton through
the radiation of electroweak gauge bosons ($W$ and $Z$) in Equation
\ref{eq:6}.

\begin{eqnarray}
\varGamma\left(l^{*}\rightarrow Vl\right)=\frac{1}{8}\frac{g_{V}^{2}}{4\pi}f_{V}^{2}\frac{m_{l^{*}}^{3}}{\Lambda^{2}}\left(1-\frac{m_{V}^{2}}{m_{l^{*}}^{2}}\right)^{2}\left(2+\frac{m_{V}^{2}}{m_{l^{*}}^{2}}\right)\label{eq:6}
\end{eqnarray}

Here the symbol $V$ is used $W^{-}$, $W^{+}$, and $Z$ bosons.
Hence $m_{V}$ represent the mass of $W^{-}$, $W^{+}$, and $Z$
bosons, and $g_{V}$ represent $g_{W}=e/sin\theta_{W}$ and $g_{Z}=e/cos\theta_{W}$,
respectively. The constants $f_{\gamma}$ and $f_{V}$ described in
Equation \ref{eq:7}. 

\begin{eqnarray*}
f_{\gamma}=fT_{3}+f^{'}\frac{Y}{2}
\end{eqnarray*}

\begin{eqnarray}
f_{Z}=fT_{3}cos^{2}\theta_{W}-f^{'}\frac{Y}{2}sin^{2}\theta_{W}\label{eq:7}
\end{eqnarray}

\begin{eqnarray*}
f_{W}=\frac{f}{\sqrt{2}}
\end{eqnarray*}

Here $T_{3}$ is the third component of weak isospin. $T_{3}=-1/2$
and the weak hypercharge $Y=-1$ for excited leptons. The coefficients
$f$ and $f^{'}$ are known free parameters, but compositeness dynamics
determine their value. We take the greatest value $f=f^{'}=1$ in
this paper.

Excited leptons can also decay to the Standart Model leptons via four-fermion
contact interaction that has the following formula;

\begin{eqnarray}
\varGamma\left(l^{*}\rightarrow lf\bar{f}\right)=\frac{m_{l^{*}}}{96\pi}\left(\frac{m_{l^{*}}}{\Lambda}\right)^{4}N_{C}^{'}S^{'}\label{eq:8}
\end{eqnarray}

Where $l$ and $f$ are the Standard Model leptons and fermions, $N_{C}^{'}$
is a color factor of quantum chromodynamics and, their value is $1$
or $3$ for leptons and quarks, respectively. $S^{'}$ represent a
combinatorial factor that value is $1$ for $f\neq l$ or 2 for $f=l$.
One can see from Equation \ref{eq:5}, the partial decay width of
excited lepton through gauge interactions via photon radiation is
proportional to the third power of the excited lepton's mass. As seen
from Equation \ref{eq:6} that the partial decay width of the excited
leptons through other gauge interactions is proportional to the inverse
power of the mass of the excited lepton. As illustrated in Equation
\ref{eq:8}, the excited lepton's partial decay width through contact
interaction is proportional to the fifth power of the excited lepton's
mass. Consequently, as the mass of the excited lepton increases, the
partial decay channel through contact interactions will be more dominant
than the decay channels through gauge interactions. However, one can
see from Equations \ref{eq:5}, \ref{eq:6}, and \ref{eq:8}, the
decay channels of excited leptons are proportional to the inverse
square power of the compositeness scale through gauge interactions.
Also, the decay channels of the excited leptons through contact interactions
are proportional to the inverse fourth power of the compositeness
scale. Suppose We increase the mass of excited leptons and select the compositeness scale bigger than their mass values. The dominance of partial contact interaction decay width over the partial gauge interaction decay width is decreased.

As mentioned in the second chapter, there appears to be a renewed
interest in TeV scale center-of-mass energy muon colliders. Such a
multi-TeV scale muon collider could offer a significant opportunity
to search for the excited muon. We have calculated the decay width and
cross section for regions that fall into some of these colliders'
energy ranges for the excited muon.

\subsection*{3.1 6 TeV Muon Collider}

\begin{figure}[h!]
\includegraphics[scale=0.6]{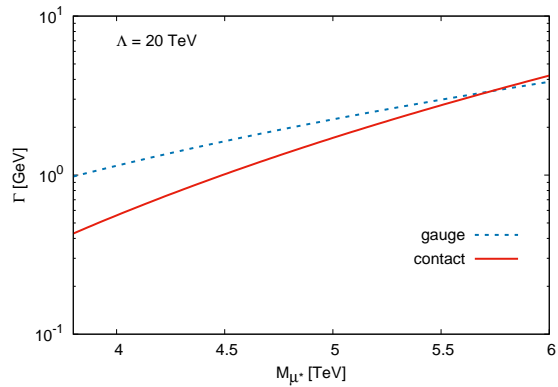}

\caption{\label{fig1:Total-decay-widths}The total decay widths of the excited muon through the gauge and contact interaction for $\Lambda=20$ TeV.}
\end{figure}

\begin{figure}[h!]
\includegraphics[scale=0.6]{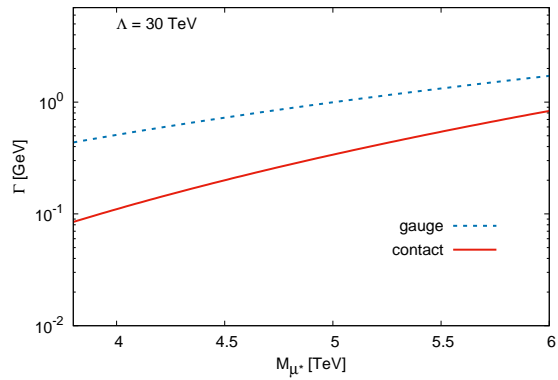}

\caption{\label{fig2:Total-decay-widths}The total decay widths of the excited muon through the gauge and contact interaction with $\Lambda=30$ TeV.}
\end{figure}

\begin{figure}[h!]
\includegraphics[scale=0.6]{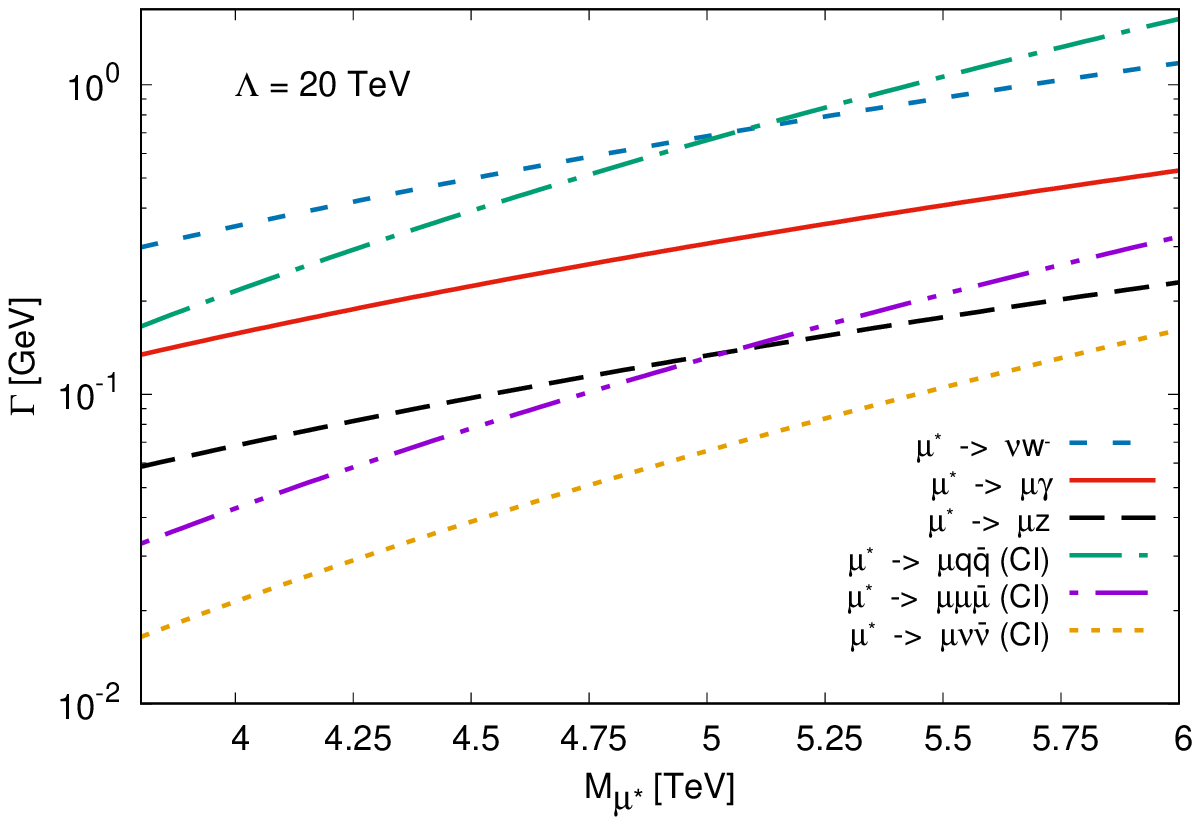}

\caption{\label{fig1:Partial-decay-width}Partial decay widths of the excited muon
via the contact and the gauge interactions with $\Lambda=20$ TeV.}
\end{figure}

\begin{figure}[h!]
\includegraphics[scale=0.6]{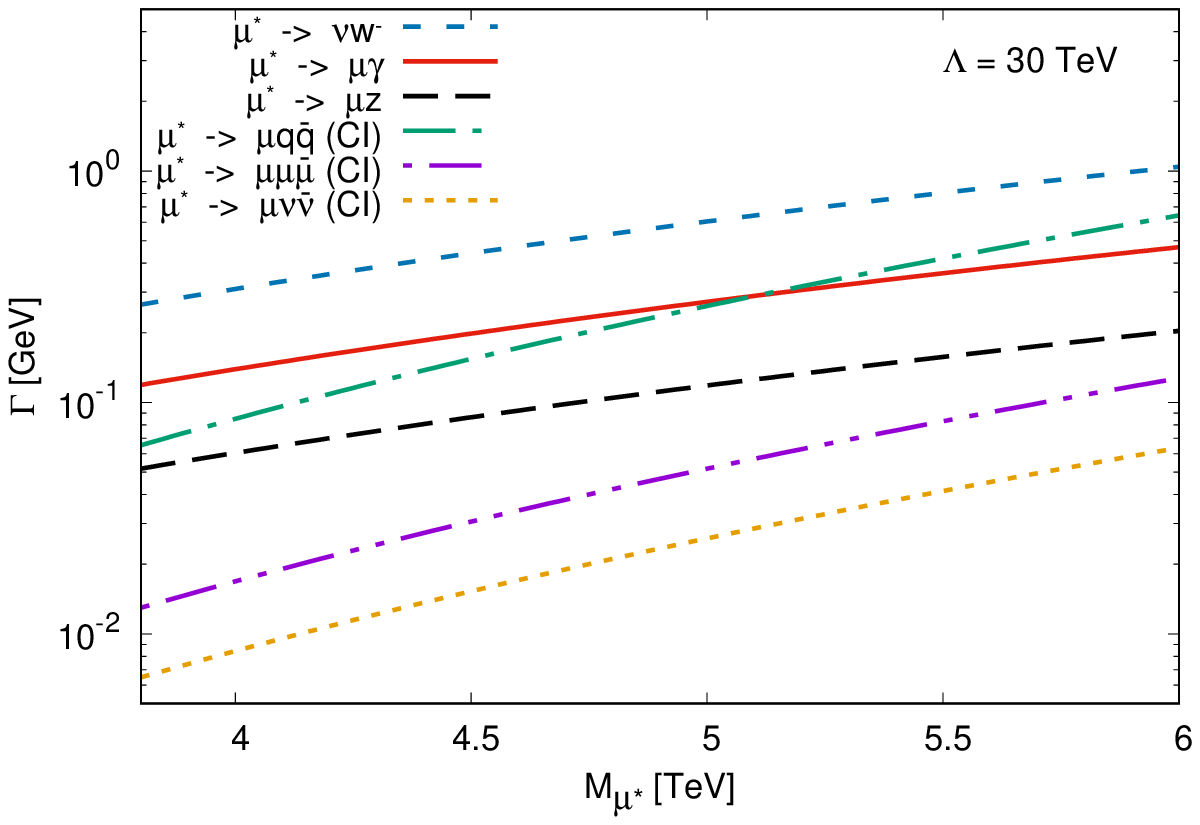}

\caption{\label{fig2:Partial-decay-width}Partial decay width of the excited muon
via the contact and the gauge interactions with $\Lambda=30$ TeV.}
\end{figure}

We implemented contact and gauge interaction lagrangians of the excited
leptons in Equation \ref{eq:2} and Equation \ref{eq:4} into the
CalcHEP simulation package \citep{belyaev2013calchep} via the LanHEP
program \citep{semenov1997automatic,semenov2016lanhep}. Then we calculated
the total decay width of the excited muon for the contact and the
gauge interactions and plotted Figure \ref{fig1:Total-decay-widths}
and Figure \ref{fig2:Total-decay-widths}. As seen in Figure \ref{fig1:Total-decay-widths}
that the gauge interaction mode dominates the contact interaction
mode up to $5.7$ TeV for $\Lambda=20$ TeV. If we take the compositeness
scale $\Lambda=30$ TeV, it can be seen from Figure \ref{fig2:Total-decay-widths}
that the decay width of the excited muon via gauge interactions is
completely more dominant than the decay width via contact interactions.
We presented the partial decay widths of the excited muon in Figure
\ref{fig1:Partial-decay-width} and Figure \ref{fig2:Partial-decay-width}
. As seen in Figure \ref{fig1:Partial-decay-width} and Figure \ref{fig2:Partial-decay-width},
$\mu^{*}\rightarrow\nu W^{-}$, $\mu^{*}\rightarrow\mu^{-}q\bar{q}$,
and $\mu^{*}\rightarrow\mu^{-}\gamma$ decay channels have larger
decay width values than other decay channels. Here $q$ symbol represents
$q=u,d,s,c,b,t$ the quarks.

\begin{figure}[h!]
\includegraphics[scale=0.4]{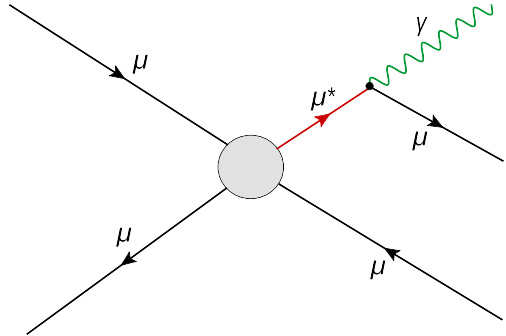}

\caption{\label{fig5:Feynman-diagram-for}Feynman diagram for single production
of an excited muon via the contact interaction at a muon collider. The
excited muon decays via electromagnetic gauge interaction to a photon
and a muon.}
\end{figure}

\begin{figure}[h!]
\includegraphics[scale=0.6]{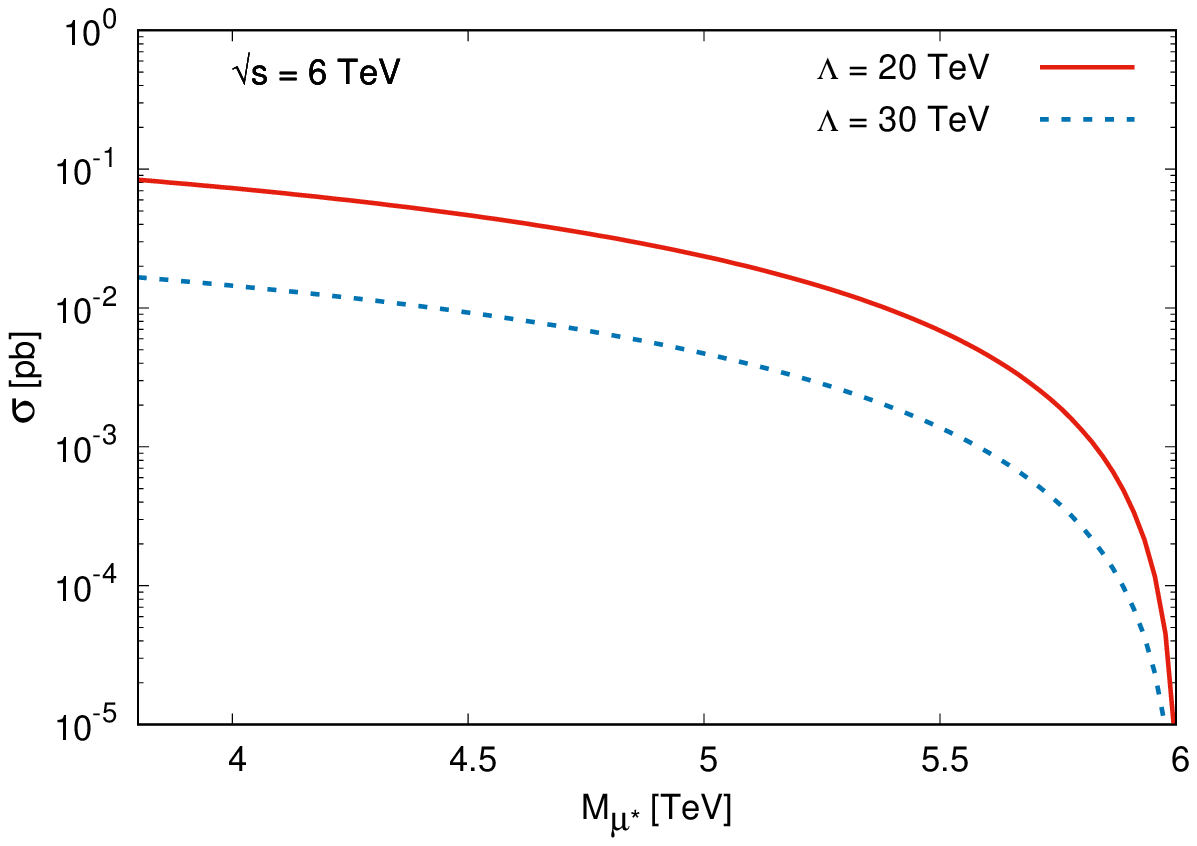}

\caption{\label{fig6:Total-cross-section}Total cross section for single production
of the excited muon via contact interaction at muon collider with $\sqrt{s}=6$
TeV.}
\end{figure}

Thus, We examined the production of the excited muon via contact interaction
and decays to photon and muon via gauge interaction at TeV scale muon
colliders in this paper. Our signal process is $\mu^{-}\mu^{+}\rightarrow\mu^{*}\mu^{+}\rightarrow\gamma\mu^{-}\mu^{+}$.
We depicted the Feynman diagram of the signal process in Figure \ref{fig5:Feynman-diagram-for}.
Figure \ref{fig5:Feynman-diagram-for} represents the single production
of the excited muon via contact interaction. In here, the excited muon decays
to a photon and a muon via gauge interaction at TeV scale muon colliders.
We calculated the total cross sections for the single production of
the excited muon signal process $\mu^{-}\mu^{+}\rightarrow\mu^{*}\mu^{+}\rightarrow\gamma\mu^{-}\mu^{+}$
at muon collider with $\sqrt{s}=6$ TeV. Then we presented the total
cross sections for single production of excited muons with $\Lambda=20$
and $30$ TeV in Figure \ref{fig6:Total-cross-section}. The compositeness
scale value of the excited muon is taken as a higher value, as shown in Figure \ref{fig6:Total-cross-section}, that single production
cross section values of the excited muon decrease.

\begin{figure}[h!]
	\includegraphics[scale=0.6]{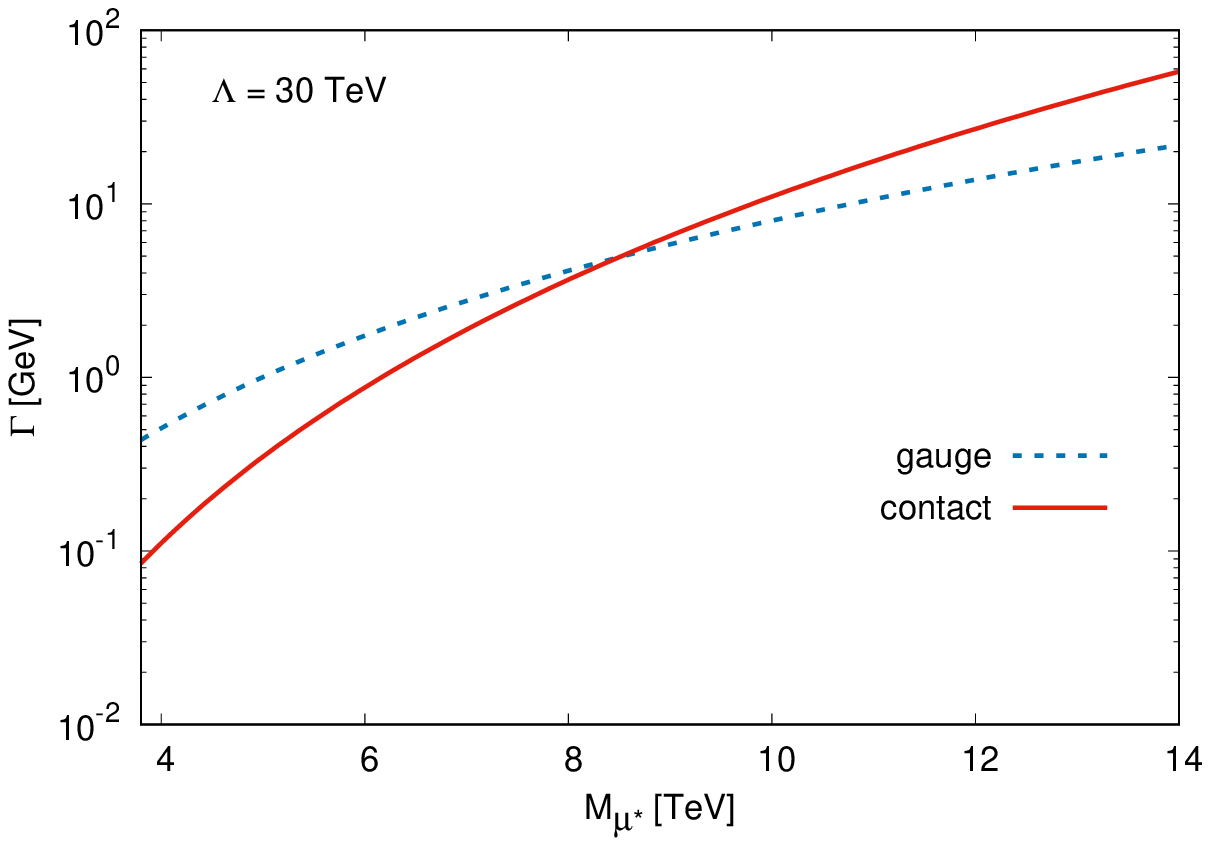}
	
	\caption{\label{fig7:Total-decay-width}The total decay width of the excited muon for the contact and the gauge interactions with $\Lambda=30$ TeV.}
\end{figure}

\subsection*{3.2 14 TeV Muon Collider}

\begin{figure}[h!]
\includegraphics[scale=0.6]{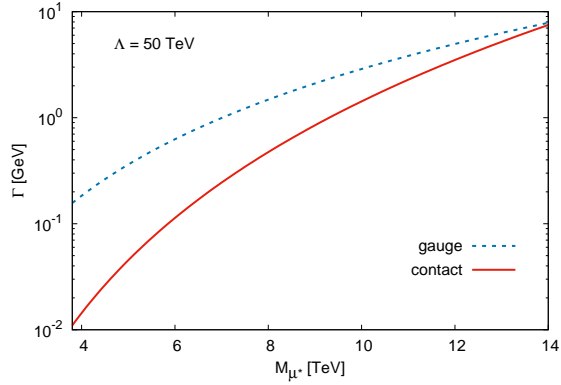}

\caption{\label{fig8:Total-decay-width}The total decay width of the excited muon
for the contact and the gauge interactions with $\Lambda=50$ TeV.}
\end{figure}

\begin{figure}[h!]
\includegraphics[scale=0.6]{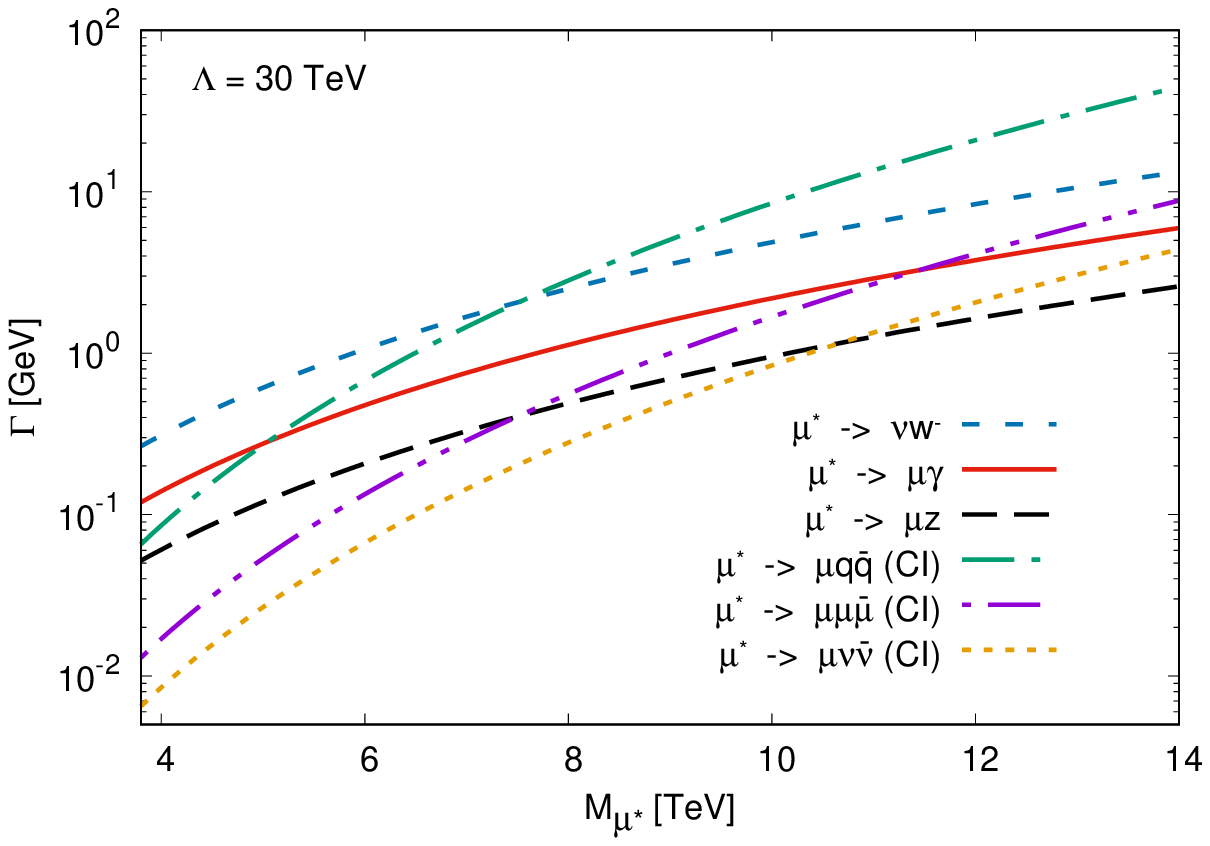}

\caption{\label{fig9:Partial-decay-width}The partial decay width of excited muon via the contact and the gauge interactions with $\Lambda=30$ TeV.}
\end{figure}

\begin{figure}[h!]
\includegraphics[scale=0.6]{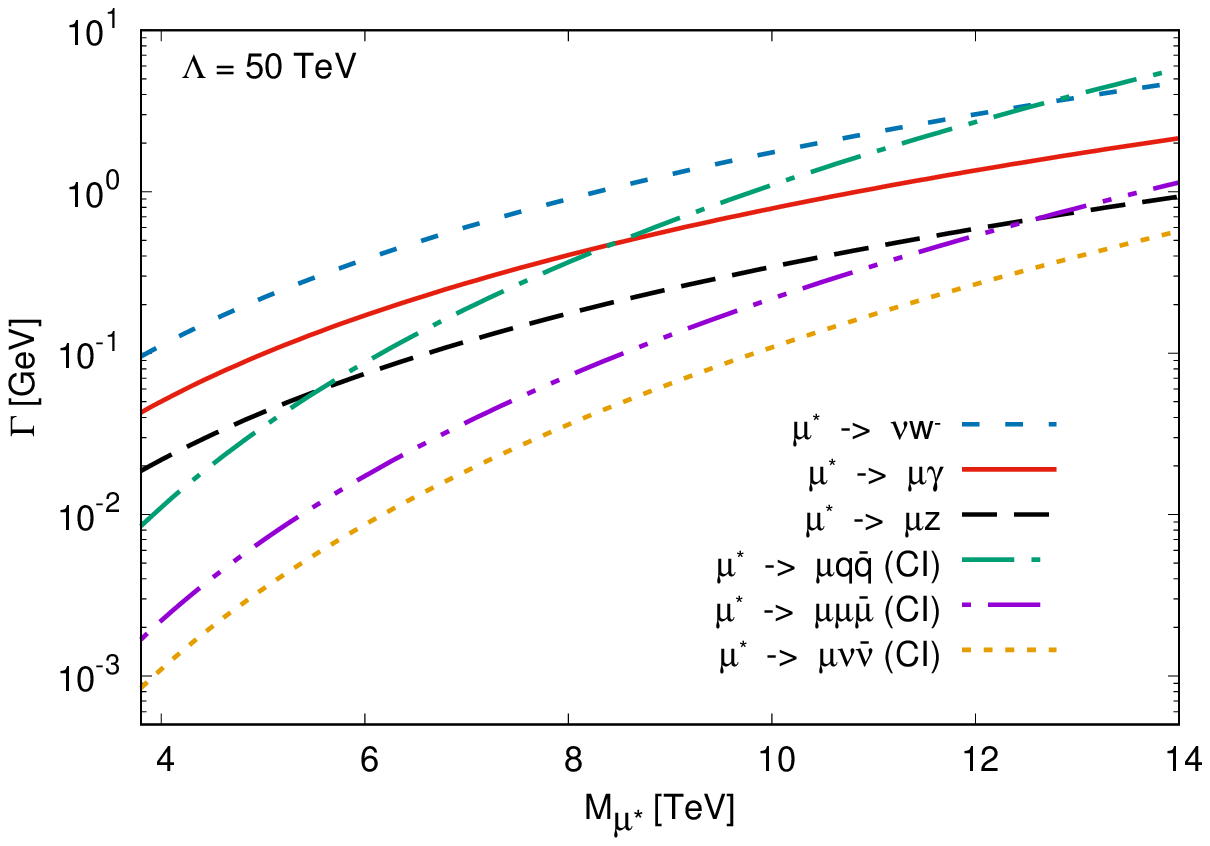}

\caption{\label{fig10:Partial-decay-width}The partial decay width of excited muon via the contact and the gauge interactions with $\Lambda=50$ TeV.}
\end{figure}

\begin{figure}[h!]
\includegraphics[scale=0.6]{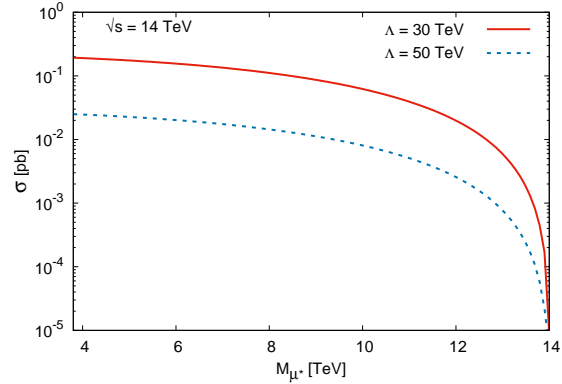}

\caption{\label{fig11:Total-cross-section}The total cross section for single production of the excited muon via contact interaction at muon collider with $\sqrt{s}=14$ TeV.}
\end{figure}

As we mentioned in the previous subsection, we added the contact and
gauge interaction Lagrangians of the excited leptons to the CalcHEP package
program \citep{belyaev2013calchep} using the LanHEP program \citep{semenov1997automatic,semenov2016lanhep}.
Then we used the CalcHEP package program and calculated the total
decay width of the excited muon for the contact and the gauge interactions
and plotted Figure \ref{fig7:Total-decay-width} and Figure \ref{fig8:Total-decay-width}.
One can see from Figure \ref{fig7:Total-decay-width} that the gauge
interaction mode dominates the contact interaction mode up to $8.4$
TeV for $\Lambda=20$ TeV. If we take the compositeness scale $\Lambda=50$
TeV, it can be seen from Figure \ref{fig8:Total-decay-width} that
the decay width of the excited muon via gauge interactions is more
dominant than the decay width via contact interactions. We introduced
the partial decay widths of the excited muon in Figure \ref{fig9:Partial-decay-width}
and Figure \ref{fig10:Partial-decay-width}. One can see from Figure
\ref{fig9:Partial-decay-width} and Figure \ref{fig10:Partial-decay-width}
, $\mu^{*}\rightarrow\nu W^{-}$, $\mu^{*}\rightarrow\mu^{-}q\bar{q}$,
and $\mu^{*}\rightarrow\mu^{-}\gamma$ decay channels have larger
decay width values than other decay channels. Here $q$ symbol represents
$q=u,d,s,c,b,t$ the quarks. As mentioned in the previous sub-section,
we chose as $\mu^{-}\mu^{+}\rightarrow\mu^{*}\mu^{+}\rightarrow\gamma\mu^{-}\mu^{+}$
the signal process for this reason. The total cross sections for single
production via contact interaction and decays to photon and muon via
gauge interaction of excited muon with $\Lambda=30$ and $\Lambda=50$
TeV at the center of mass-energy is $14$ TeV muon collider is shown
in Figure \ref{fig11:Total-cross-section}. The compositeness scale
value of the excited muon is taken as a higher value, as it is seen from
Figure \ref{fig11:Total-cross-section} that single production cross
section values of the excited muon decrease.

\subsection*{3.3 100 TeV Muon Collider}

\begin{figure}[h!]
\includegraphics[scale=0.6]{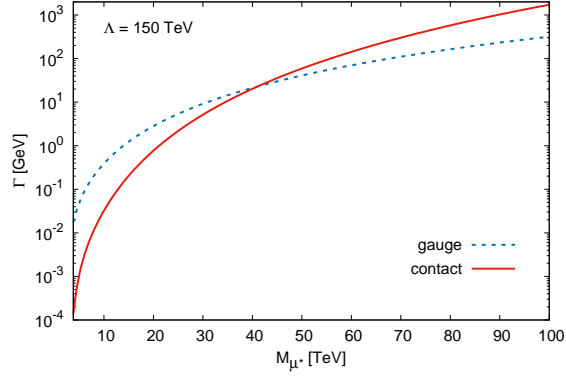}

\caption{\label{fig12:Total-decay-width}The total decay width of the excited muon for the contact and the gauge interactions with $\Lambda=150$ TeV.}

\end{figure}

\begin{figure}[h!]
\includegraphics[scale=0.6]{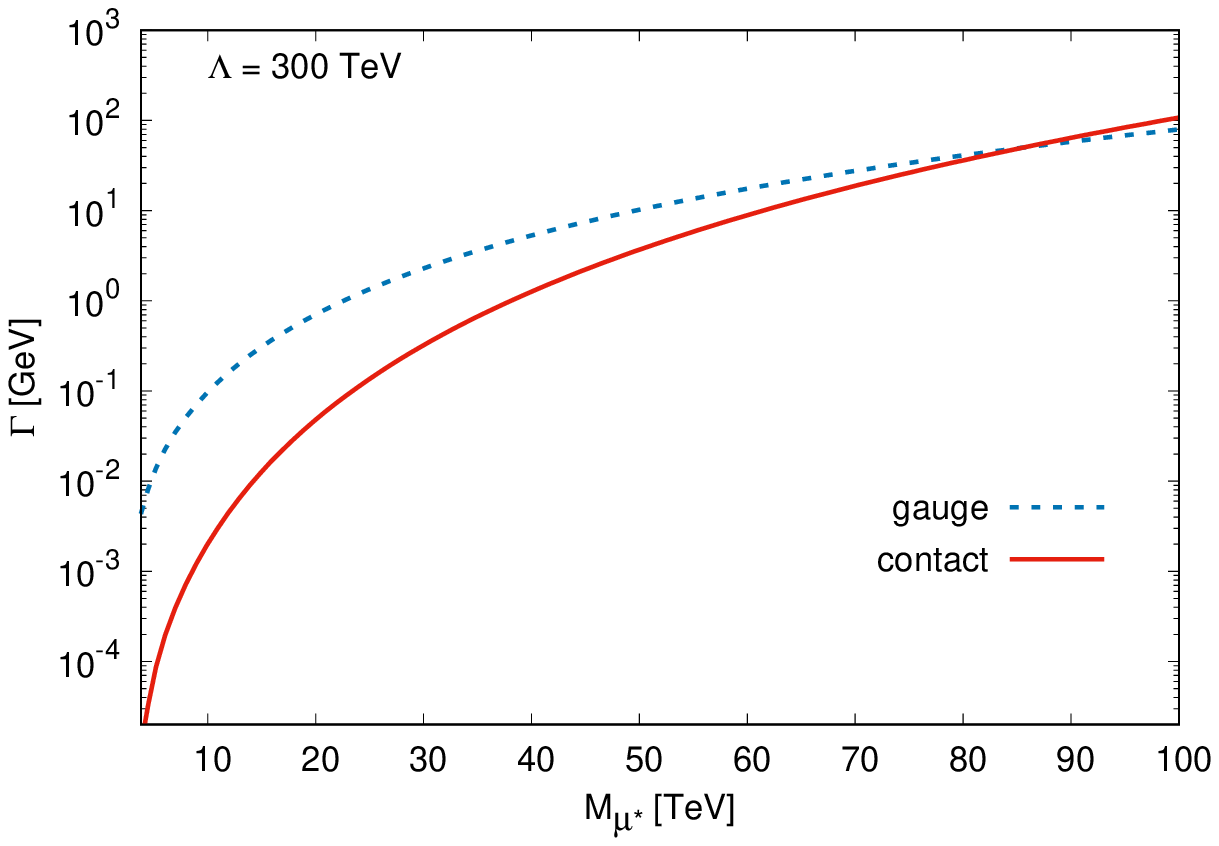}

\caption{\label{fig13:Total-decay-width}The total decay width of the excited muon for the contact and the gauge interactions with $\Lambda=300$ TeV.}

\end{figure}

\begin{figure}[h!]
\includegraphics[scale=0.6]{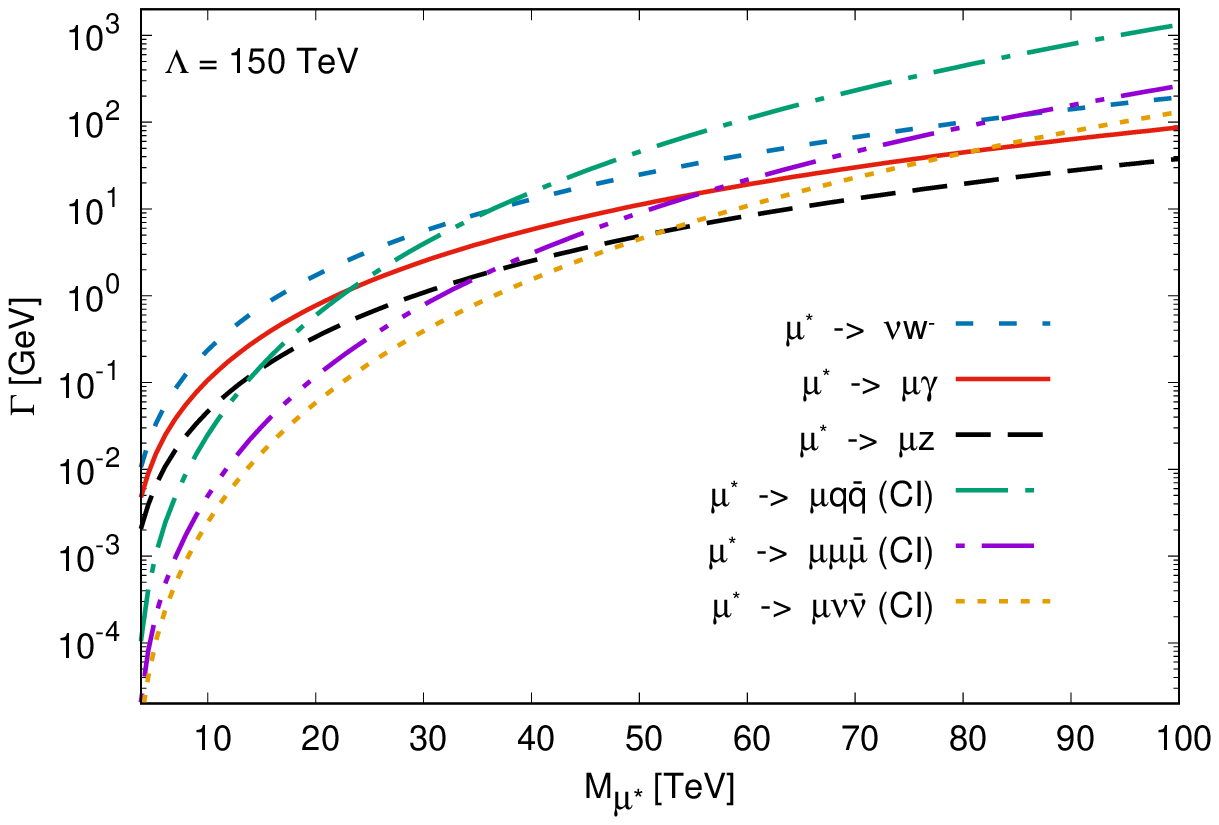}

\caption{\label{fig14:Partial-decay-width}The partial decay width of excited muon via the contact and the gauge interactions with $\Lambda=150$ TeV.}

\end{figure}

\begin{figure}[h!]
\includegraphics[scale=0.6]{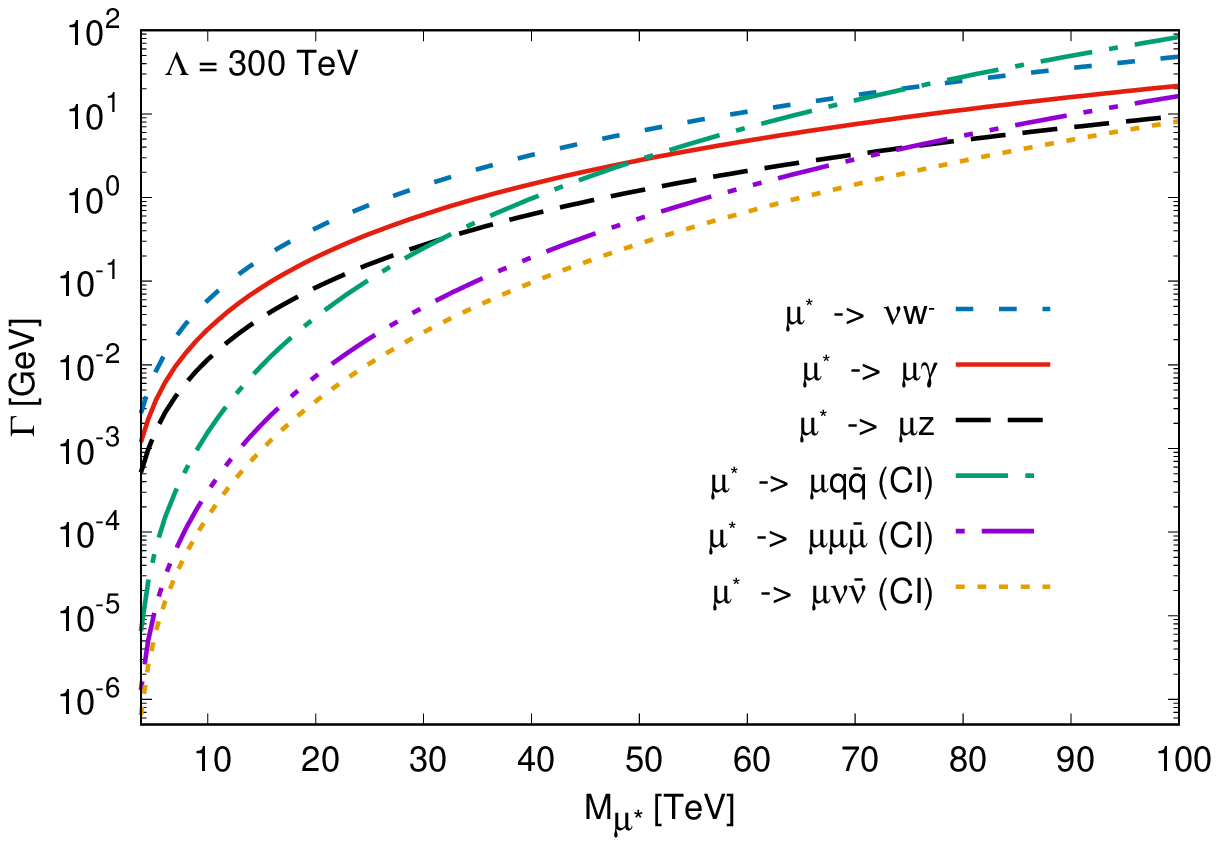}

\caption{\label{fig15:Partial-decay-width}The partial decay width of excited muon via the contact and the gauge interactions with $\Lambda=300$ TeV.}

\end{figure}

\begin{figure}[h!]
\includegraphics[scale=0.6]{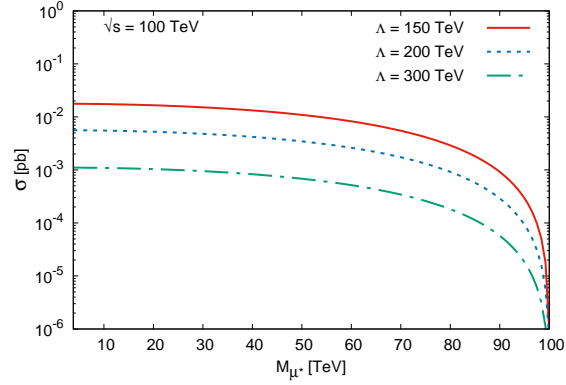}

\caption{\label{fig16:Total-cross-section}The total cross section for single production of the excited muon via the contact interaction at muon collider with $\sqrt{s}=100$
TeV.}

\end{figure}

We computed the total decay width of the excited muon for the contact
and the gauge interactions. We plotted Figure \ref{fig12:Total-decay-width}
and Figure \ref{fig13:Total-decay-width}. One can observe from Figure
\ref{fig12:Total-decay-width} that the gauge interaction mode dominates
the contact interaction mode up to $43$ TeV for $\Lambda=150$ TeV.
If we take the compositeness scale $\Lambda=300$ TeV, it can be noticed
from Figure \ref{fig13:Total-decay-width} that the decay width of
the excited muon via the gauge interactions is more dominant than the
decay width via the contact interactions. We presented the partial decay
widths of the excited muon in Figure \ref{fig14:Partial-decay-width}
and Figure \ref{fig15:Partial-decay-width}. As seen from Figure \ref{fig14:Partial-decay-width},
and Figure \ref{fig15:Partial-decay-width}, $\mu^{*}\rightarrow\nu W^{-}$,
$\mu^{*}\rightarrow\mu^{-}q\bar{q}$, and $\mu^{*}\rightarrow\mu^{-}\gamma$
decay channels have larger decay width values than other decay channels.
Here $q$ symbol represents $q=u,d,s,c,b,t$ the quarks. For this
reason, $\mu^{-}\mu^{+}\rightarrow\mu^{*}\mu^{+}\rightarrow\gamma\mu^{-}\mu^{+}$
process has been chosen as the signal process as in the previous subsection.
Then we calculated the total cross-section for single production via
the contact interaction and decays to photon and muon via the gauge interaction of the excited muon with $\Lambda=150$, $\Lambda=200$, and $\Lambda=300$
TeV at the center-of-mass energy is $\sqrt{s}=100$ TeV muon collider.
After that, we plotted the total cross section for the excited muon's
single production as in Figure \ref{fig16:Total-cross-section}. The
compositeness scale value of the excited muon is taken as a higher value,
as seen from Figure \ref{fig16:Total-cross-section}, single
production cross section values of the excited muon decrease. 

\section*{4. SIGNAL AND BACKGOUND ANALYSIS}

As addressed in the previous section, the $\mu^{*}\rightarrow\nu W^{-}$,
$\mu^{*}\rightarrow\mu^{-}q\bar{q}$, and $\mu^{*}\rightarrow\mu^{-}\gamma$
partial decay channels of the excited muon are more dominant than
the other decay channels. Consequently, we chose $\mu^{-}\mu^{+}\rightarrow\mu^{*}\mu^{+}\rightarrow\gamma\mu^{-}\mu^{+}$
as our signal process. According to this process, excited muon is
produced at muon colliders through the contact interaction. Later, it decays
into photon and muon via the gauge interactions.
In this study, we could also choose the $\mu^{-}\mu^{+}\rightarrow\mu^{*}\mu^{+}\rightarrow q\bar{q}\mu^{-}\mu^{+}$ process
for the signal process in which the excited muon is produced through
the contact interactions and decay via the contact interactions. Here $q$
symbol represents $q=u,d,s,c,b,t$ the quarks. However, in this case,
the number of final-state particles would be higher than the $\mu^{-}\mu^{+}\rightarrow\mu^{*}\mu^{+}\rightarrow\gamma\mu^{-}\mu^{+}$
process. For this reason, the background process's cross section
would be more since the number of particles in the final state would
be higher. So it would be more challenging to separate the signal from the background. Hence, we chose $\mu^{-}\mu^{+}\rightarrow\mu^{*}\mu^{+}\rightarrow\gamma\mu^{-}\mu^{+}$
as the signal process. The background process that corresponds to
this signal process is $\mu^{-}\mu^{+}\rightarrow\gamma\mu^{-}\mu^{+}$.
As mentioned in the previous section, we added lagrangians to the CalcHEP simulation program via the LanHEP program. Those lagrangians representing excited leptons' interactions with Standard Model fermions via the contact and the gauge interactions. We applied $p_{T}^{\gamma}>25$ GeV, $p_{T}^{\mu^{-}}>25$
GeV, $p_{T}^{\mu^{+}}>25$ GeV cut values for photon, muon and anti-muon
of the final state particles in the signal and the background processes.
Here $p_{T}$ represents the transverse momentum of the final state
particles. Also, we applied the $\Delta R(\gamma\mu^{-})>0.7$, and
$\Delta R(\gamma\mu^{+})>0.7$ cut value to separate the final state
leptons from photons. Here $\Delta R$ is separation cuts, and $\Delta R=\sqrt{\Delta\eta^{2}+\Delta\phi^{2}}$. 

\subsection*{4.1 Final State Particle Distributions and Significance Calculus
at 6 TeV Muon Collider}

\begin{figure}[h!]
\includegraphics[scale=0.6]{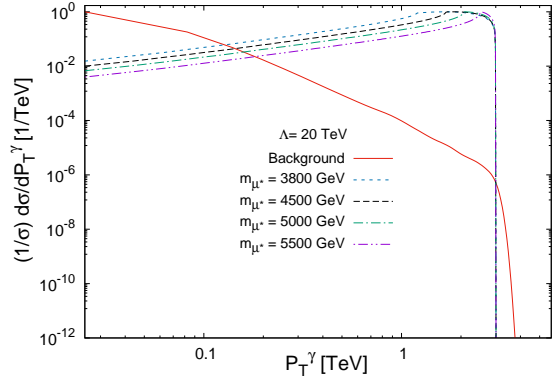}

\caption{\label{fig17:The-transverse-momentum}The transverse momentum distribution
of final state photon for the signal and the background processes
with $\Lambda=20$ TeV.}

\end{figure}

\begin{figure}[h!]
\includegraphics[scale=0.6]{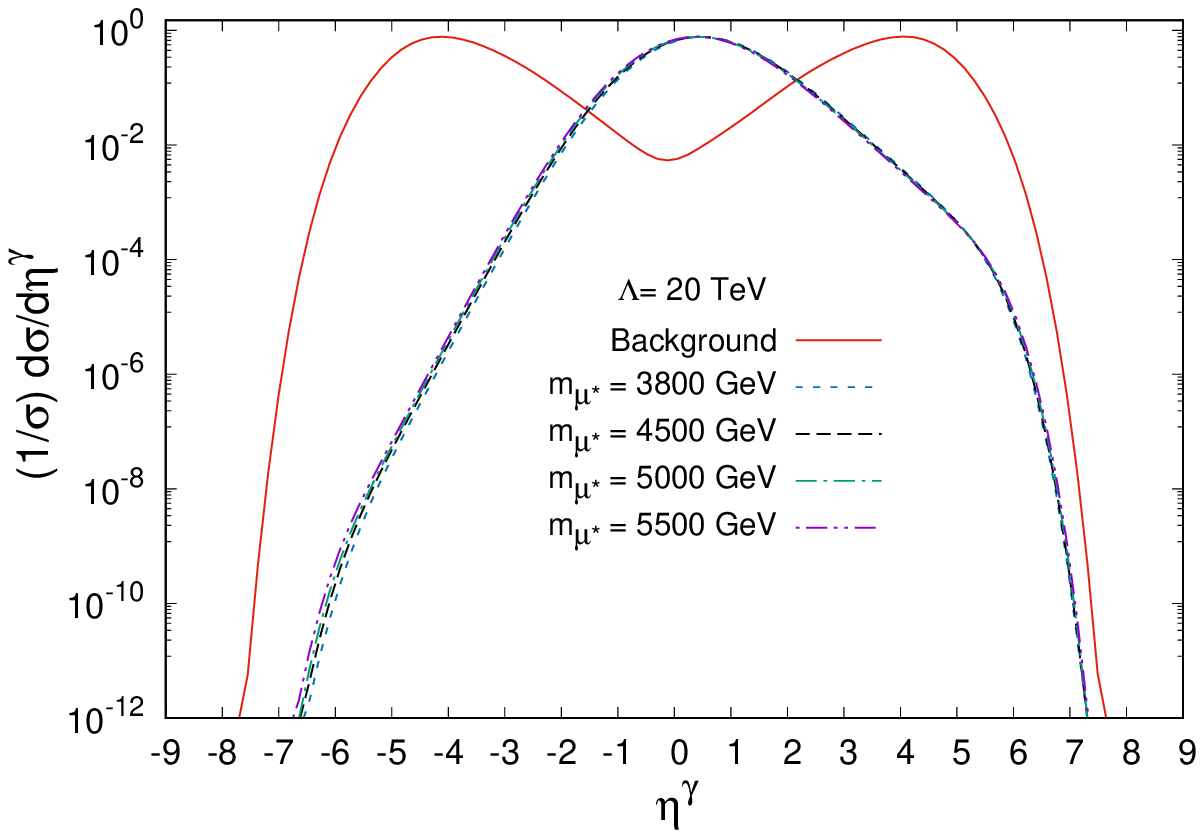}

\caption{\label{fig18:The-pseudorapidity-distribution}The pseudorapidity distribution
of final state photon for the signal and the background processes
with $\Lambda=20$ TeV.}
\end{figure}

\begin{figure}[h!]
\includegraphics[scale=0.6]{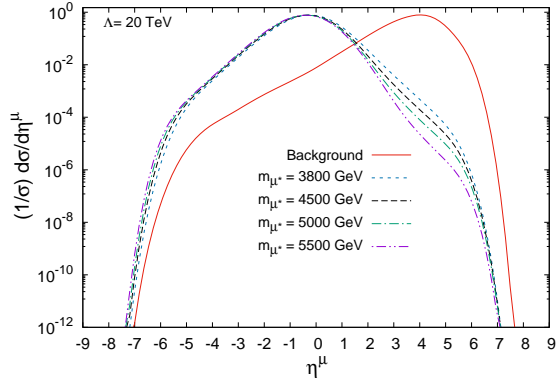}

\caption{\label{fig19:The-pseudorapidity-distribution}The pseudorapidity distribution
of final state muon for the signal and the background processes with
$\Lambda=20$ TeV.}

\end{figure}

In the present subsection, signal and background analysis are performed
for the excited muon at 6 TeV center-of-mass energy collider predicted
in the MAP and the LEMMA programs. Since the cross section of the background process is larger than the signal process's cross section, it is not possible to identify the signal from the background with the acceptance cuts mentioned above. We needed the transverse momentum and the pseudorapidity
distributions of the final state particles in the signal and background
processes to determine the cuts to identify the signal over the
background. It is observed that excited muon decay to photon and muon
in the signal process. Since the muon and photon distributions are
similar, we exhibited the photon's transverse momentum and pseudorapidity
distributions for the illustration. As shown in Figure \ref{fig17:The-transverse-momentum},
when the cut of $p_{T}^{\gamma}>400$ GeV is applied, this cut leaves
the cross section value of the signal process almost unchanged. At the same time, it dramatically reduces the cross section value of the background
process. Since the muon's final state distribution is similar to the
photon's final state distribution, the same cut value $p_{T}^{\mu^{-}}>400$
GeV can be used for the muon. We plotted pseudorapidity distributions
of final state photon and muon for the signal and the background process
in Figures \ref{fig18:The-pseudorapidity-distribution} and \ref{fig19:The-pseudorapidity-distribution}
. As shown in Figure \ref{fig18:The-pseudorapidity-distribution},
when the $-2.5<\eta^{\gamma}<2.5$ pseudorapidity cut interval is
applied to the final state photon, the cross-section value of the
signal remains almost unchanged. The background cross section value decreases.
When Figure \ref{fig19:The-pseudorapidity-distribution} is examined,
during the $-4<\eta^{\mu^{-}}<1.5$ pseudorapidity cut interval is
applied, the signal's cross-section value remains almost unchanged,
while the background cross-section value is significantly reduced.
For all these reasons, we applied $p_{T}^{\gamma}>400$ GeV, $-2.5<\eta^{\gamma}<2.5$
cut values for the final state photon in the signal and the background
processes. For the final state muon, we applied $p_{T}^{\mu^{-}}>400$
GeV, $-4<\eta^{\mu^{-}}<1.5$ cut values. We preferred to apply the
cut values $p_{T}^{\mu^{+}}>400$ GeV, $-2.5<\eta^{\mu^{+}}<2.5$
for the final state anti-muon in the signal and the background processes.
Besides, $|M_{\gamma\mu^{-}}-M_{\mu^{*}}|<500$ GeV mass window cut is another
essential cut used to differentiate the signal from the background.
We presented the cuts we determined as a result of all these inquiries
in Equation \ref{eq9:}.

\begin{eqnarray*}
p_{T}^{\gamma,\mu^{-},\mu^{+}}>400 & \:GeV, & -2.5<\eta^{\gamma,\mu^{+}}<2.5,
\end{eqnarray*}

\begin{eqnarray}
-4<\eta^{\mu^{-}}<1.5, & \:\Delta R_{(\mu^{-}\gamma),(\mu^{+}\gamma)}>0.7\label{eq9:}
\end{eqnarray}

\begin{eqnarray*}
 & |M_{\gamma\mu^{-}}-M_{\mu^{*}}|<500 & GeV
\end{eqnarray*}

We used Equation \ref{eq10:} for statistical significance calculation.
Here $N_{s}$ represents the number of signal events and $N_{s}=\sigma_{s}\sqrt{L_{int}}$.
$N_{B}$ in the Equation \ref{eq10:} represents the number of background
events, and $N_{B}=\sigma_{B}L_{int}$. Also, the $\sigma_{s}$ represents
the signal cross section, $\sigma_{B}$ represents the background
cross section, and $L_{int}$ represents the integrated luminosity.

\begin{eqnarray}
S=\frac{N_{s}}{\sqrt{N_{s}+N_{B}}}\label{eq10:}
\end{eqnarray}

We used cut sets in the Equation \ref{eq9:} and the integrated Luminosity
values listed for the $6$ TeV center-of-mass energy of the muon collider
in Table 1. Then, we have performed discovery ($5\sigma$), observation
($3\sigma$), and exclusion ($2\sigma$) analysis for the mass of
the excited muon. Table \ref{tab:table2} shows excited muon mass
limits for the discovery ($5\sigma$), the observation ($3\sigma$), and the exclusion ($2\sigma$) at the muon collider with 6 TeV center-of-mass energy. 

\begin{table*}[h!]
\caption{\label{tab:table2}The attainable mass limit for the excited muon at
$6$ TeV muon collider.}

\begin{ruledtabular}
\begin{tabular}{lcccccc}
\multicolumn{7}{c}{$\sqrt{S}=6$ TeV}\tabularnewline
$L_{int}$ & \multicolumn{3}{c}{$510\:fb^{-1}$} & \multicolumn{3}{c}{$1200\:fb^{-1}$}\tabularnewline
Significance & $5\sigma$ & $3\sigma$ & $2\sigma$ & $5\sigma$ & $3\sigma$ & $2\sigma$\tabularnewline
$\Lambda=20$ TeV & $5.59$ TeV & $5.60$ TeV & $5.63$ TeV & $5.60$ TeV & $5.62$ TeV & $5.66$ TeV\tabularnewline
\end{tabular}
\end{ruledtabular}

\end{table*}

\begin{figure}[h!]
\includegraphics[scale=0.6]{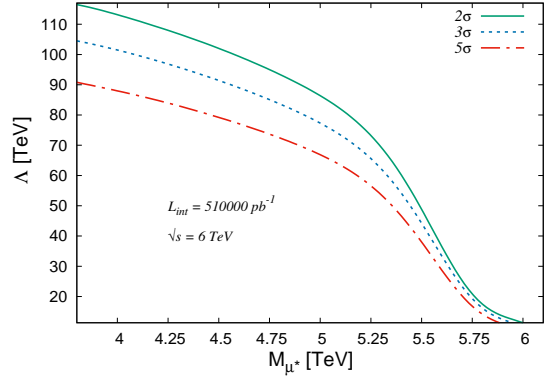}

\caption{\label{fig20:The-compositeness-scale}The compositeness scale versus
the mass plots for the excited muon at $6$ TeV center-of-mass energy
muon collider with $L_{int}=510\,fb^{-1}$.}
\end{figure}

\begin{figure}[h!]
\includegraphics[scale=0.6]{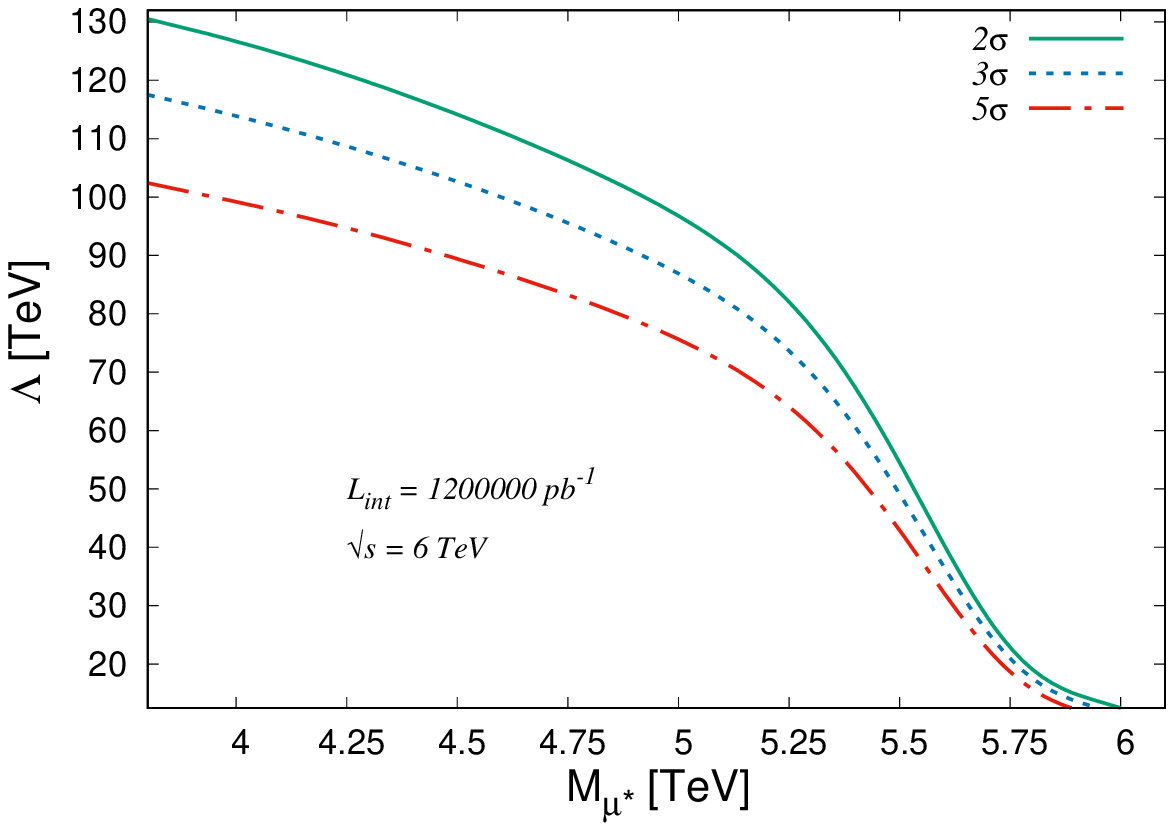}

\caption{\label{fig21:The-compositenes-scale}The compositenes scale versus
the mass plots for the excited muon at $6$ TeV center-of-mass energy
muon collider with $L_{int}=1200\,fb^{-1}$.}
\end{figure}

So far, we have taken the compositeness scale as $\Lambda=20$ TeV
in the calculations was performed. However, it is essential to determine
the excited muon's compositeness scale. Hence we scanned it by taking
different values for the compositeness scale. Then we calculated the
attainable compositeness scale limits for the excited muon at the muon
collider with $\sqrt{S}=6$ TeV and $L_{int}=510\,fb^{-1}$. We plotted
Figure \ref{fig20:The-compositeness-scale} for the muon collider with
$6$ TeV center-of-mass energy. Figure \ref{fig20:The-compositeness-scale}
represents the achievable compositeness scale limits versus the excited
muon mass plot. As shown in the Figure \ref{fig20:The-compositeness-scale}
, the excited muon with mass $M_{\mu^{*}}=4$ TeV can be discovered
up to $\Lambda=88.1$ TeV compositeness scale value for $L_{int}=510\,fb^{-1}$
at the muon collider with $6$ TeV center-of-mass energy. Similar calculations
were made within $L_{int}=1200\,fb^{-1}$ total integrated luminosity
value. Then we plotted Figure \ref{fig21:The-compositenes-scale}
. Figure \ref{fig21:The-compositenes-scale} represents the attainable
compositeness scale limits corresponding to the excited muon mass
values plot. As illustrated in Figure \ref{fig21:The-compositenes-scale},
the excited muon with mass $M_{\mu^{*}}=4$ TeV can be discovered
up to $99.2$ TeV compositeness scale value at the muon collider with
$6$ TeV center of mass-energy and $L_{int}=1200\,fb^{-1}$. Detailed
limit values regarding the achievable compositeness scale of the excited
muon for the excited muon mass are listed in Table \ref{tab3:Attainable-compositeness-scale}. 

\begin{table}[h!]
\caption{\label{tab3:Attainable-compositeness-scale}The attainable compositeness scale limits for some excited muon mass values at muon collider with 6 TeV center-of-mass energy.}

\begin{ruledtabular}
\begin{tabular}{lcccccc}
\multicolumn{7}{c}{$\sqrt{S}=6$ TeV}\tabularnewline
$L_{int}$ & \multicolumn{3}{c}{$510\:fb^{-1}$} & \multicolumn{3}{c}{$1200\:fb^{-1}$}\tabularnewline
 & \multicolumn{3}{c}{Attainable Compositeness Scale ($\Lambda$)} & \multicolumn{3}{c}{Attainable Compositeness Scale ($\Lambda$)}\tabularnewline
Significance & $5\sigma$ & $3\sigma$ & $2\sigma$ & $5\sigma$ & $3\sigma$ & $2\sigma$\tabularnewline
$M_{\mu^{*}}=3.8$ TeV & $90.80$ TeV & $104.5$ TeV & $116.5$ TeV & $102.4$ TeV & $117.5$ TeV & $130.5$ TeV\tabularnewline
$M_{\mu^{*}}=4.0$ TeV & $88.10$ TeV & $101.6$ TeV & $113.3$ TeV & $99.20$ TeV & $113.9$ TeV & $126.8$ TeV\tabularnewline
$M_{\mu^{*}}=4.2$ TeV & $85.00$ TeV & $98.10$ TeV & $109.2$ TeV & $95.90$ TeV & $110.1$ TeV & $122.4$ TeV\tabularnewline
$M_{\mu^{*}}=4.4$ TeV & $81.45$ TeV & $94.00$ TeV & $104.8$ TeV & $91.90$ TeV & $105.4$ TeV & $117.4$ TeV\tabularnewline
$M_{\mu^{*}}=4.6$ TeV & $77.45$ TeV & $89.40$ TeV & $99.90$ TeV & $87.45$ TeV & $100.5$ TeV & $111.5$ TeV\tabularnewline
$M_{\mu^{*}}=4.8$ TeV & $72.95$ TeV & $84.30$ TeV & $93.90$ TeV & $82.35$ TeV & $94.50$ TeV & $105.3$ TeV\tabularnewline
$M_{\mu^{*}}=5.0$ TeV & $67.70$ TeV & $78.20$ TeV & $87.50$ TeV & $76.60$ TeV & $87.80$ TeV & $98.00$ TeV\tabularnewline
$M_{\mu^{*}}=5.2$ TeV & $61.90$ TeV & $71.50$ TeV & $80.20$ TeV & $70.00$ TeV & $80.50$ TeV & $89.80$ TeV\tabularnewline
$M_{\mu^{*}}=5.4$ TeV & $53.70$ TeV & $62.40$ TeV & $69.90$ TeV & $61.05$ TeV & $70.50$ TeV & $78.60$ TeV\tabularnewline
$M_{\mu^{*}}=5.6$ TeV & $18.17$ TeV & $20.20$ TeV & $21.85$ TeV & $19.88$ TeV & $21.90$ TeV & $23.60$ TeV\tabularnewline
\end{tabular}
\end{ruledtabular}

\end{table}

As shown in Figures \ref{fig20:The-compositeness-scale}, \ref{fig21:The-compositenes-scale},
and Table \ref{tab3:Attainable-compositeness-scale}, the muon collider
with $6$ TeV center-of-mass energy will give an excellent potential
search for a compositeness scale. Even this collider gives opportunity up to $100$ TeV compositeness scale value can be achieved. 

\subsection*{4.2 Final State Particle Distributions and Significance Calculus
at 14 TeV Muon Collider}

\begin{figure}[h!]
\includegraphics[scale=0.6]{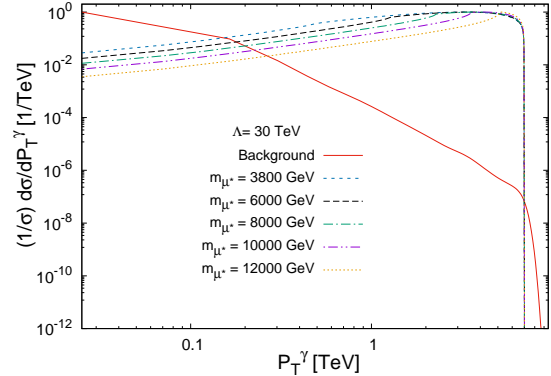}

\caption{\label{fig22:The-transverse-momentum}The transverse momentum distribution
of final state photon for the signal and the background processes
with $\Lambda=30$ TeV.}

\end{figure}

\begin{figure}[h!]
\includegraphics[scale=0.6]{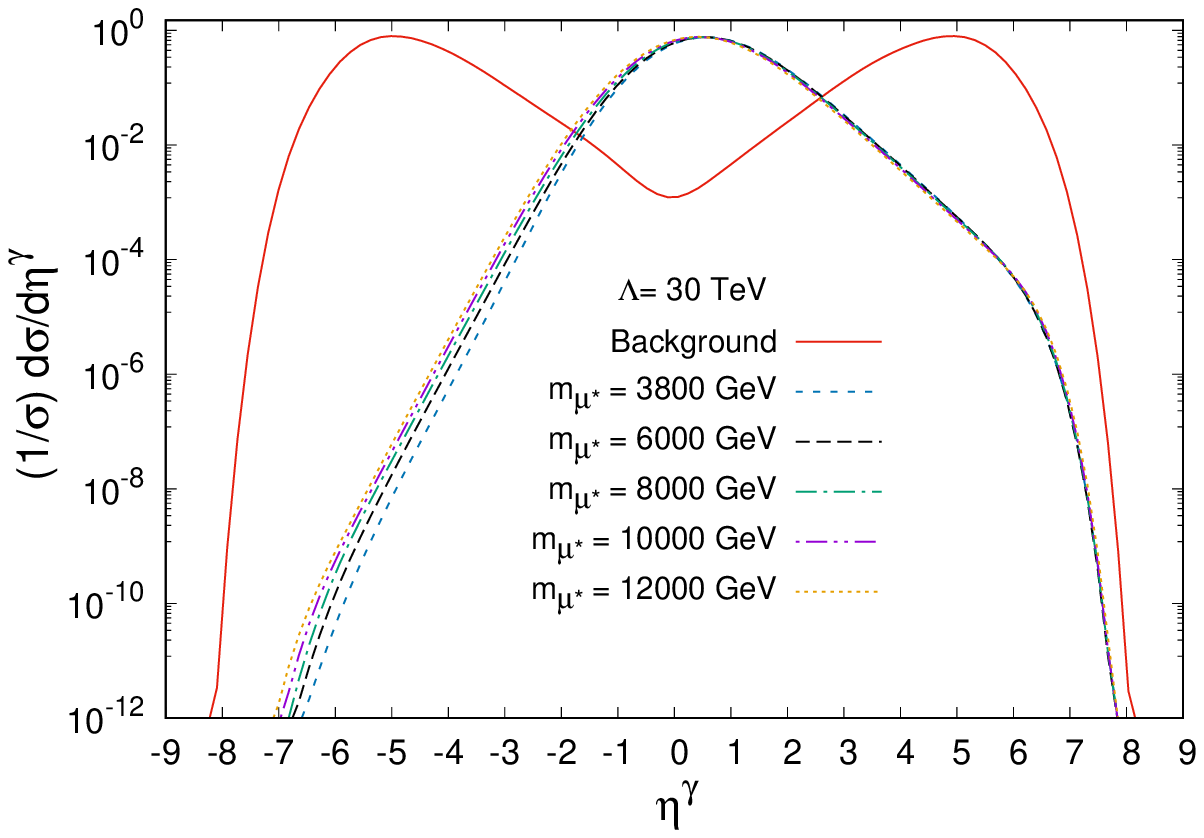}

\caption{\label{fig23:The-pseudorapidity-distribution}The pseudorapidity distribution
of final state photon for the signal and the background processes
with $\Lambda=30$ TeV}

\end{figure}

\begin{figure}[h!]
\includegraphics[scale=0.6]{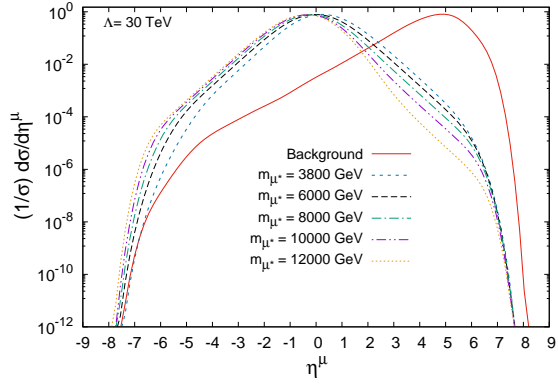}

\caption{\label{fig24:The-pseudorapidity-distribution}The pseudorapidity distribution
of final state muon for the signal and the background processes with
$\Lambda=30$ TeV}

\end{figure}

We conducted a signal and background analysis for the excited muon at the muon collider with $\sqrt{s}=14$ TeV. It will be achieved using the LHC collider ring as a muon collider at the CERN. As illustrated
at the beginning of chapter 4, we implemented lagrangians, which describe
the gauge and the contact interactions of the excited leptons with the SM fermions, to the CalcHEP program through the LanHeP program. Our signal process
is $\mu^{-}\mu^{+}\rightarrow\mu^{*}\mu^{+}\rightarrow\gamma\mu^{-}\mu^{+}$
and our background process is $\mu^{-}\mu^{+}\rightarrow\gamma\mu^{-}\mu^{+}$.
For signal and background analysis, we applied generator level $p_{T}^{\gamma}>25$
GeV, $p_{T}^{\mu^{-}}>25$ GeV, $p_{T}^{\mu^{+}}>25$ GeV cut values
for photon, muon and anti-muon of the final state particles in the
signal and the background processes. Also, we applied the $\Delta R(\gamma\mu^{-})>0.7$,
and $\Delta R(\gamma\mu^{+})>0.7$ cut value to separate the final
state leptons from the photons. After that, we plotted the transverse
momentum and the pseudorapidity distributions of the final state particles
in the signal and the background processes to determine the cuts to identify the signal from the background. Since the final state of the muon
and the photon distributions are similar, we exhibited the photon's transverse
momentum and pseudorapidity distributions for the illustration. Figure
\ref{fig22:The-transverse-momentum} shows the transverse momentum
distributions of photons in the signal and the background processes. We took
the compositeness scale as $\Lambda=30$ TeV, and the excited muon mass
values are $M_{\mu^{*}}=3800,6000,8000,10000$, and $12000$ GeV in
the final state particle distribution calculations. As shown in Figure
\ref{fig22:The-transverse-momentum}, when we applied the cut $p_{T}^{\gamma}>500$
GeV for the final state photon, this cut leaves the cross-section
value of the signal process almost unchanged. At the same time, it dramatically reduces the cross-section value of the background process. Since the muon's final state distribution is similar to the photon's final state
distribution, the same cut value $p_{T}^{\mu^{-}}>500$ GeV can be
used for the muon. We plotted pseudorapidity distributions of final
state photon and muon for the signal and the background process in Figures
\ref{fig23:The-pseudorapidity-distribution} and \ref{fig24:The-pseudorapidity-distribution}.
As illustrated in Figure \ref{fig23:The-pseudorapidity-distribution},
when the $-3<\eta^{\gamma}<3$ pseudorapidity cut interval is applied
to the final state photon, the cross-section value of the signal remains
almost unchanged. The background cross-section value decreases. When Figure
\ref{fig24:The-pseudorapidity-distribution} is examined, during the
$-4<\eta^{\mu^{-}}<2.5$ pseudorapidity cut interval is applied, the
signal's cross-section value remains almost unchanged, while the background
cross-section value is significantly reduced. For all these reasons,
we applied $p_{T}^{\gamma}>500GeV$, $-3<\eta^{\gamma}<3$ cut values
for the final state photon in the signal and the background processes.
For the final state muon, we applied $p_{T}^{\mu^{-}}>500GeV$, $-4<\eta^{\mu^{-}}<2.5$
cut values. Also, we preferred to apply the cut values $p_{T}^{\mu^{+}}>500GeV$,
$-2.5<\eta^{\mu^{+}}<2.5$ for the final state anti-muon in the signal
and the background processes. Besides, $|M_{\gamma\mu^{-}}-M_{\mu^{*}}|<500$ GeV mass window cut is another essential cut used to identify the signal
from the background. We summarized all of these cuts we determined
to use in signal and background analysis at the muon collider with 14
TeV center-of-mass energy in the Equation \ref{eq:11}.

\medskip

\begin{eqnarray*}
p_{T}^{\gamma,\mu^{-},\mu^{+}}>500 & GeV, & -3<\eta^{\gamma}<3,
\end{eqnarray*}

\begin{eqnarray}
-4<\eta^{\mu^{-}}<2.5, & -2.5<\eta^{\mu^{+}}<2.5 & ,\label{eq:11}
\end{eqnarray}

\begin{eqnarray*}
\Delta R_{(\mu^{-}\gamma),(\mu^{+}\gamma)}>0.7, & |M_{\gamma\mu^{-}}-M_{\mu^{*}}|<500 & GeV
\end{eqnarray*}

Using the statistical significance formula in the Equation \ref{eq10:}
and all cuts summarized in the Equation \ref{eq:11}, we calculated the
achievable mass limits for excited muons at the muon collider with 14
TeV center-of-mass energy. Then, we listed discovery ($5\sigma$),
observation ($3\sigma$), and exclusion ($2\sigma$) limits for the
mass of the excited muon in Table \ref{tab4:The-attainable-mass}.
As can be seen from Table \ref{tab4:The-attainable-mass}, for $L_{int}=12\,fb^{-1}$,
the excited muon can be discovered up to $13.26$ TeV at the muon
collider with $14$ TeV center-of-mass energy. High Luminosity Large
Hadron Collider (HL-LHC) can not reach this limit value even in its
entire working lifetime. From this, it can be said that the muon colliders
will be a very effective collider.

\begin{table}[h!]

\caption{\label{tab4:The-attainable-mass}The attainable mass limit for the excited muon at $14$ TeV muon collider.}

\begin{ruledtabular}
\begin{tabular}{lcccccc}
\multicolumn{7}{c}{$\sqrt{S}=14$ TeV}\tabularnewline
$L_{int}$ & \multicolumn{3}{c}{$2.4\:fb^{-1}$} & \multicolumn{3}{c}{$12\:fb^{-1}$}\tabularnewline
Significance & $5\sigma$ & $3\sigma$ & $2\sigma$ & $5\sigma$ & $3\sigma$ & $2\sigma$\tabularnewline
$\Lambda=30$ TeV & $12.52$ TeV & $13.08$ TeV & $13.31$ TeV & $13.26$ TeV & $13.43$ TeV & $13.48$ TeV\tabularnewline
\end{tabular}
\end{ruledtabular}

\end{table}

\begin{figure}[h!]
\includegraphics[scale=0.6]{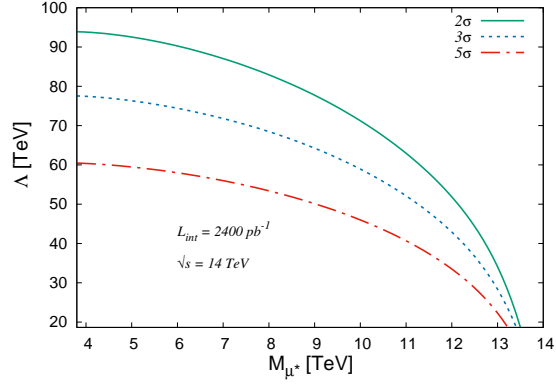}

\caption{\label{fig25:The-compositeness-scale}The compositeness scale versus
the mass plots for the excited muon at $14$ TeV center-of-mass energy
muon collider with $L_{int}=2.4\,fb^{-1}$.}

\end{figure}

\begin{figure}[h!]
\includegraphics[scale=0.6]{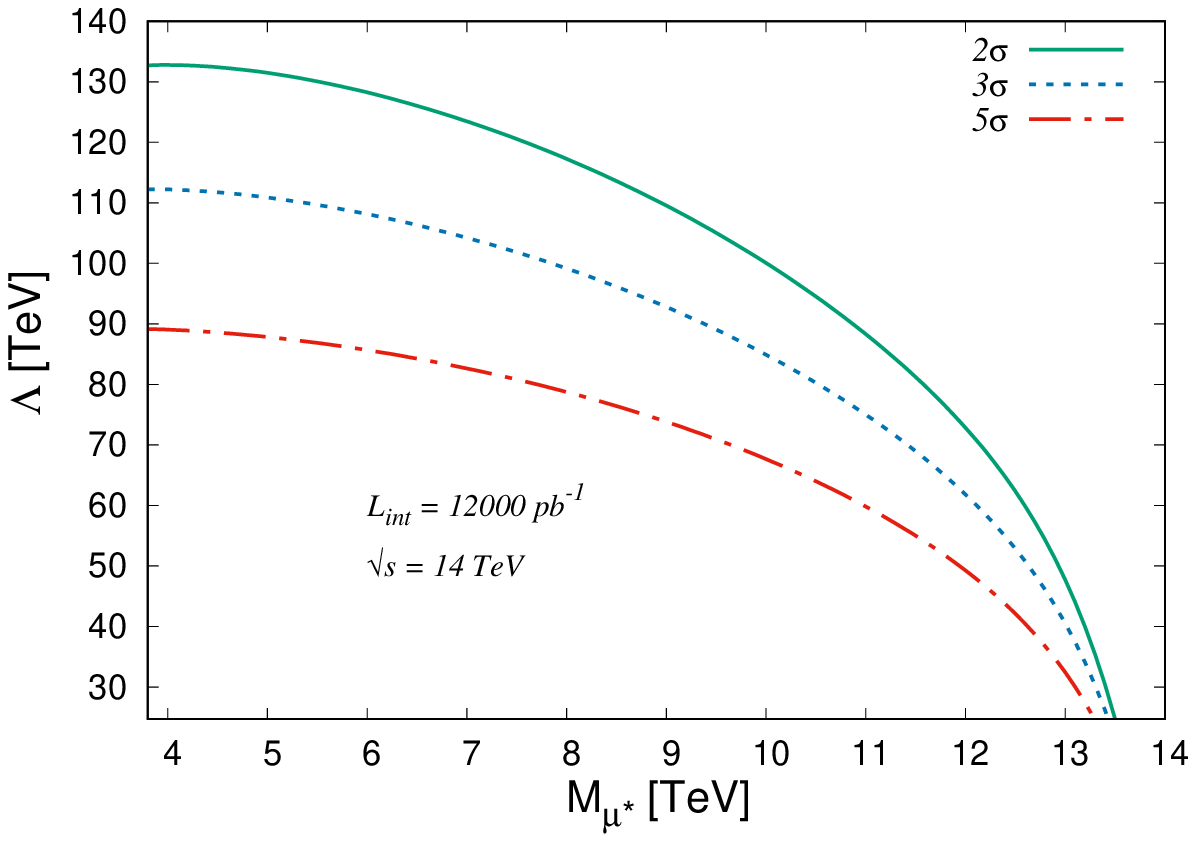}

\caption{\label{fig26:The-compositeness-scale}The compositeness scale versus
the mass plots for the excited muon at $14$ TeV center-of-mass energy
muon collider with $L_{int}=12\,fb^{-1}$.}

\end{figure}

Up to now, we performed all calculations for $\Lambda=30$ TeV in
this sub-section. However, the compositeness scale of the excited muon
may have values other than $30$ TeV. Therefore, we did a compositeness
scale scan for each different excited muon mass. Figure \ref{fig25:The-compositeness-scale}
represents the compositeness scale plot for $L_{int}=2.4\,fb^{-1}$
depending on the mass of the excited muon. As depicted in Figure \ref{fig25:The-compositeness-scale},
the excited muon with a mass value of $3.8$ TeV will be discovered
at the muon collider up to $\Lambda=60.45$ TeV compositeness scale
value. Figure \ref{fig26:The-compositeness-scale} represents the
compositeness scale plot for $L_{int}=12\,fb^{-1}$ depending on the
mass of the excited muon. As seen in Figure \ref{fig26:The-compositeness-scale},
the excited muon with a mass value of $3.8$ TeV will be discovered up
to a compositeness scale value of $\Lambda=89.14$ TeV. The detailed
compositeness scale limits obtained depending on the mass of the excited
muon are listed in Table \ref{tab5:Attainable-compositeness-scale}.
According to Table \ref{tab5:Attainable-compositeness-scale}, the
excited muon with $10$ TeV mass will be discovered up to a compositeness
scale of approximately $69$ TeV at the muon collider with $14$ TeV center-of-mass energy. The results show that the $14$ TeV center-of-mass energy muon collider will be a good alternative to Hadron-Hadron colliders
to investigate the excited muon, although it has low integrated luminosity
values.

\begin{table}[h!]

\caption{\label{tab5:Attainable-compositeness-scale}The attainable compositeness scale limits for some excited muon mass values at the muon collider with $14$ TeV center-of-mass energy.}

\begin{ruledtabular}
\begin{tabular}{lcccccc}
\multicolumn{7}{c}{$\sqrt{S}=14$ TeV}\tabularnewline
$L_{int}$ & \multicolumn{3}{c}{$2.4\:fb^{-1}$} & \multicolumn{3}{c}{$12\:fb^{-1}$}\tabularnewline
 & \multicolumn{3}{c}{Attainable Compositeness Scale ($\Lambda$)} & \multicolumn{3}{c}{Attainable Compositeness Scale ($\Lambda$)}\tabularnewline
Significance & $5\sigma$ & $3\sigma$ & $2\sigma$ & $5\sigma$ & $3\sigma$ & $2\sigma$\tabularnewline
$M_{\mu^{*}}=3.8$ TeV & $60.45$ TeV & $77.54$ TeV & $93.86$ TeV & $89.14$ TeV & $112.2$ TeV & $132.7$ TeV\tabularnewline
$M_{\mu^{*}}=5.0$ TeV & $59.65$ TeV & $76.58$ TeV & $92.88$ TeV & $88.15$ TeV & $111.3$ TeV & $132.1$ TeV\tabularnewline
$M_{\mu^{*}}=6.0$ TeV & $58.25$ TeV & $74.78$ TeV & $90.68$ TeV & $86.07$ TeV & $108.7$ TeV & $128.9$ TeV\tabularnewline
$M_{\mu^{*}}=7.0$ TeV & $56.32$ TeV & $72.27$ TeV & $87.60$ TeV & $83.15$ TeV & $104.9$ TeV & $124.3$ TeV\tabularnewline
$M_{\mu^{*}}=8.0$ TeV & $53.82$ TeV & $69.03$ TeV & $83.58$ TeV & $79.36$ TeV & $99.93$ TeV & $118.2$ TeV\tabularnewline
$M_{\mu^{*}}=9.0$ TeV & $50.65$ TeV & $64.92$ TeV & $78.53$ TeV & $74.59$ TeV & $93.75$ TeV & $110.7$ TeV\tabularnewline
$M_{\mu^{*}}=10$ TeV & $46.66$ TeV & $59.77$ TeV & $72.20$ TeV & $68.62$ TeV & $86.05$ TeV & $101.3$ TeV\tabularnewline
$M_{\mu^{*}}=11$ TeV & $41.53$ TeV & $53.17$ TeV & $64.18$ TeV & $61.00$ TeV & $76.38$ TeV & $89.82$ TeV\tabularnewline
$M_{\mu^{*}}=12$ TeV & $34.68$ TeV & $44.44$ TeV & $53.70$ TeV & $51.03$ TeV & $63.99$ TeV & $73.36$ TeV\tabularnewline
$M_{\mu^{*}}=13$ TeV & $24.40$ TeV & $31.39$ TeV & $38.12$ TeV & $36.16$ TeV & $45.77$ TeV & $54.40$ TeV\tabularnewline
\end{tabular}
\end{ruledtabular}

\end{table}

\subsection*{4.3 Final State Particle Distributions and Significance Calculus
at $100$ TeV Muon Collider}

\begin{figure}[h!]
\includegraphics[scale=0.6]{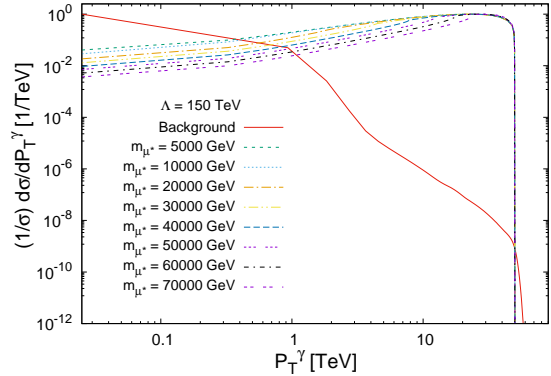}

\caption{\label{fig27:The-transverse-momentum}The transverse momentum distribution
of final state photon for the signal and the background processes
with $\Lambda=150$ TeV.}

\end{figure}

In this subsection, we performed a signal background analysis for
the excited muon at the 100 TeV center-of-mass energy muon collider,
which will be obtained using the FCC collider ring as a muon collider
at the CERN. As we mentioned at the beginning of chapter 4, our signal
process is $\mu^{-}\mu^{+}\rightarrow\mu^{*}\mu^{+}\rightarrow\gamma\mu^{-}\mu^{+}$,
and the background process is $\mu^{-}\mu^{+}\rightarrow\gamma\mu^{-}\mu^{+}$.
Firstly, we put acceptance cuts $p_{T}^{\gamma}>25$ GeV, $p_{T}^{\mu^{-}}>25$
GeV, $p_{T}^{\mu^{+}}>25$ GeV on muon, photon and, anti-muon, respectively.
In addition to these acceptance cuts, we used $-2.5<\eta^{\mu^{+}}<2.5$
pseudorapidity interval for anti-muon and $\Delta R(\gamma\mu^{-})>0.7$,
and $\Delta R(\gamma\mu^{+})>0.7$ acceptance cuts for the separation
photon from muon and anti-muon. Since these cuts are not
sufficient to identify the signal from the background, we plotted
the transverse momentum and the pseudorapidity distributions of the photon,
muon, and anti-muon, which are the final state particles of the signal
and the background processes. We performed calculations for $\Lambda=150$
TeV and $M_{\mu^{*}}=5000$, $10000$, $20000$, $30000$, $40000$,
$50000$, $60000$, and $70000$ GeV in this sub-section. Figure \ref{fig27:The-transverse-momentum}
represents the transverse momentum distributions of the final state
photons in the signal and the background processes. When we applied the
cut $p_{T}^{\gamma}>500$ GeV for the final state photon, this cut
leaves the cross-section value of the signal process almost unchanged,
while it dramatically reduces the cross-section value of the background
process. Since the muon's final state distribution is similar to the
photon's final state distribution, the same cut value $p_{T}^{\mu^{-}}>500$
GeV can be used for the muon. We plotted pseudorapidity distributions
of final state photon and muon for the signal and the background process
in Figures 28 and 29. As illustrated in Figure 28, while the $-3<\eta^{\gamma}<3.5$
pseudorapidity cut interval is applied to the final state photon,
the cross-section value of the signal remains almost unchanged the background
cross-section value decreases. When Figure 29 is examined, during
the $-4<\eta^{\mu^{-}}<3.5$ pseudorapidity cut interval is applied,
the signal's cross-section value remains almost unchanged, while the
background cross-section value is significantly reduced. Because of
all these reasons, we applied $p_{T}^{\gamma}>500$ GeV, $-3<\eta^{\gamma}<3.5$
cut values for the final state photon in the signal and the background
processes. For the final state muon, we applied $p_{T}^{\mu^{-}}>500$
GeV, $-4<\eta^{\mu^{-}}<3.5$ cut values. Also, we preferred to apply
the cut values $p_{T}^{\mu^{+}}>500$ GeV, $-2.5<\eta^{\mu^{+}}<2.5$
for the final state anti-muon in the signal and the background processes.
Moreover, $|M_{\gamma\mu^{-}}-M_{\mu^{*}}|<500$ GeV mass window cut is another
essential cut used to identify the signal from the background.
We overview all of these cuts we determined to use in signal and the background analysis at the muon collider with 100 TeV center-of-mass energy in Equation 12.

\begin{figure}[h!]
\includegraphics[scale=0.6]{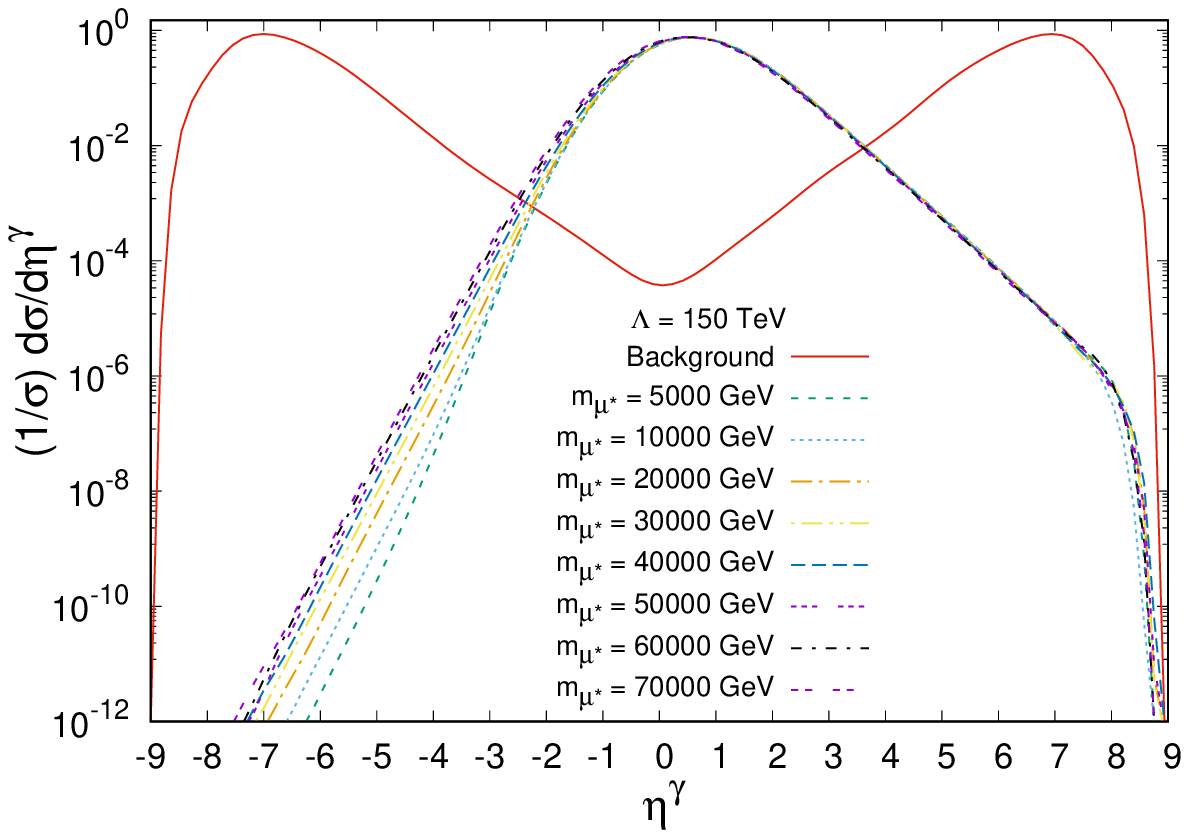}

\caption{\label{fig28:The-pseudorapidity-distribution}The pseudorapidity distribution
of final state photon for the signal and the background processes
with $\Lambda=150$ TeV}

\end{figure}

\begin{figure}[h!]
\includegraphics[scale=0.6]{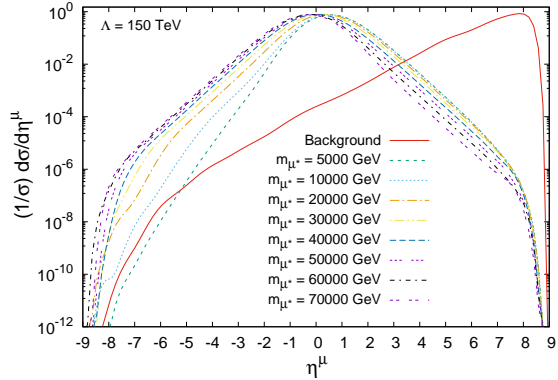}

\caption{\label{fig29:The-pseudorapidity-distribution}The pseudorapidity distribution
of final state muon for the signal and the background processes with
$\Lambda=150$ TeV.}

\end{figure}

\begin{eqnarray*}
p_{T}^{\gamma,\mu^{-},\mu^{+}}>500 & GeV, & -3<\eta^{\gamma}<3.5,
\end{eqnarray*}

\begin{eqnarray}
-4<\eta^{\mu^{-}}<3.5, & -2.5<\eta^{\mu^{+}}<2.5,\label{eq12:}
\end{eqnarray}

\begin{eqnarray*}
\Delta R_{(\mu^{-}\gamma),(\mu^{+}\gamma)}>0.7, &  & |M_{\gamma\mu^{-}}-M_{\mu^{*}}|<500GeV
\end{eqnarray*}

We used the statistical significance formula of Equation 10, cut set
in Equation 12, and an integrated luminosity value for a muon collider
with $100$ TeV center-of-mass energy in Table \ref{tab:Muon-antimuon-colliders-and}. Then, we calculated discovery ($5\sigma$), observation ($3\sigma$),
and exclusion ($2\sigma$) limits for the excited muon with $\Lambda=150$
TeV. As a result of the calculations, the excited muon will be discovered
up to $94.20$ TeV, observed up to $96.53$ TeV, and excluded up to
$97.66$ TeV at the muon collider with $100$ TeV center-of-mass energy
for $L_{int}=100\,fb^{-1}$. The fact that the excited muon will be discovered
at the muon collider up to $94.20$ TeV is a very high mass value for
the excited muon. It indicates that the muon colliders will be a very
effective collider for an excited muon. 

We did calculations for the excited muon in this sub-section; we took
the value of the compositeness scale has been taken as $\Lambda=150$
TeV up to now. However, for each mass of the excited muon, the compositeness
scale can take different values. Therefore, we calculated the upper
limits for the compositeness scale for some excited muon masses at
muon collider with the $100$ TeV center-of-mass energy. The excited muon
with a mass of $20$ TeV will be discovered up to $425.2$ TeV compositeness
scale value for $L_{int}=100\,fb^{-1}$. The excited muon with a mass
of $90$ TeV will be discovered up to $198.6$ TeV compositeness scale
value. As can be seen from these values, the muon collider with $100$
TeV center-of-mass energy and $L_{int}=100\,fb^{-1}$ will be able
to discover the excited muon even if it has very high compositeness
scale values. We listed the detailed results in table \ref{tab6:Attainable-compositeness-scale}.

\begin{table}[h!]
\caption{\label{tab6:Attainable-compositeness-scale}The attainable compositeness scale limits for some excited muon mass values at the muon collider with $100$ TeV center-of-mass energy.}

\begin{ruledtabular}
\begin{tabular}{lccc}
\multicolumn{4}{c}{$\sqrt{S}=100$ TeV with $L_{int}=100\,fb^{-1}$}\tabularnewline
 & \multicolumn{3}{c}{\begin{turn}{90}
\end{turn}}\tabularnewline
 & \multicolumn{3}{c}{Attainable Compositeness Scale ($\Lambda$)}\tabularnewline
Significance & $5\sigma$ & $3\sigma$ & $2\sigma$\tabularnewline
$M_{\mu^{*}}=20$ TeV & $425.2$ TeV & $546.9$ TeV & $665.5$ TeV\tabularnewline
$M_{\mu^{*}}=50$ TeV & $384.8$ TeV & $495.7$ TeV & $604.2$ TeV\tabularnewline
$M_{\mu^{*}}=70$ TeV & $323.6$ TeV & $417.4$ TeV & $509.3$ TeV\tabularnewline
$M_{\mu^{*}}=90$ TeV & $198.6$ TeV & $258.0$ TeV & $315.6$ TeV\tabularnewline
\end{tabular}
\end{ruledtabular}

\end{table}

\section*{5. CONCLUSION}

\begin{figure}[h!]
	\includegraphics[scale=0.6]{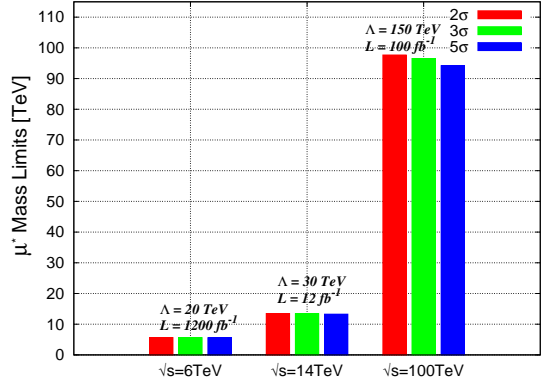}
	
	\caption{\label{fig30:Excited muon mass limits} The excited muon mass limits at various center-of-mass energy muon colliders.}
	
\end{figure}

We investigated the excited muon production via the contact interaction and
decays to the SM fermions through photon radiation at the muon colliders.
The calculations were performed at the muon colliders with $6$ TeV, $14$
TeV, and $100$ TeV center-of-mass energy options. The excited muon will
be discovered up to $5.60$ TeV mass value for $\Lambda=20$ TeV at
the muon collider with $6$ TeV center-of-mass energy, and $L_{int}=1200$$\,fb^{-1}$
integrated luminosity. It is seen that this value is close to the
center-of-mass energy value of the collider. At the muon collider
with $14$ TeV center-of-mass energy and $12\,fb^{-1}$ integrated luminosity,
the excited muon will be discovered up to $13.26$ TeV for $\Lambda=30$
TeV. This limit value is also close to the center-of-mass energy
of the collider. Since the High Luminosity Large Hadron Collider is
a hadron-hadron collider, It will not reach this mass limit for the excited
muon even during its entire run time. Moreover, the excited muon will
be discovered up to $94.20$ TeV mass value for $\Lambda=150$ TeV
at the muon collider with $100$ TeV center-of-mass energy and $L_{int}=100\,fb^{-1}$
integrated luminosity. As with other center-of-mass energy options,
this limit value is close to the collider's center-of-mass energy
value. Because the FCC proton-proton collider is a hadron-hadron collider,
it will not reach this mass limit even during its entire run time.
All these results show that the muon colliders will be a perfect collider
option for examining the excited muon. These limits on the mass of the excited muon at all center-of-mass energy options for the muon colliders are shown in Figure \ref{fig30:Excited muon mass limits}.

An excited muon with a $4$ TeV mass will be discovered at the muon collider
with a $6$ TeV center-of-mass energy and $L_{int}=1200\,fb^{-1}$
up to $99.20$ TeV compositeness scale. At the muon collider with
$14$ TeV center-of-mass energy and $L_{int}=12\,fb^{-1}$, the excited
muon with a $5$ TeV mass will be discovered up to $88.15$ TeV compositeness
scale value. Also, the excited muon with $20$ TeV mass will be discovered
up to 425.2 TeV compositeness scale at the muon collider with $100$ TeV
center-of-mass energy and $L_{int}=100\,fb^{-1}$. As can be seen
from all these data, the excited muon can be examined up to very high
compositeness scale values at the muon collider. In other words, the
muon colliders will be able to determine the compositeness scale value
of the excited muon.

All these calculations about the excited muon show that the muon colliders
will be a unique collider search for the excited muon.Since the mass of the muon is greater than the electron, it loses less energy in the acceleration process. 
Therefore, it will be easier to increase the momentum of
the muons to higher energies at accelerators. For all
these reasons, many Standard Model and Beyond Standard Model processes
will be investigated better than other types of colliders through the
muon anti-muon colliders.
\begin{acknowledgments}
Authors appreciate to the Usak University, Energy, Environment and Sustainability Application and Research Center for their help. 
\end{acknowledgments}

\bibliographystyle{apsrev4-2}
\bibliography{reference}

\end{document}